\documentstyle[prl,aps]{revtex}
\begin{document}
\title{ Hierarchical search strategy for the detection of 
gravitational waves from coalescing binaries: Extension to post-Newtonian
wave forms}
\author{S.D.Mohanty }
\address{Inter-University Centre for Astronomy and Astrophysics,
 Post Bag - 4, Ganeshkhind, Pune 411~007, India}
\maketitle 

\begin{abstract} 
The detection of gravitational waves from coalescing compact binaries
 would be a computationally
 intensive process if a single bank of template wave forms~({\em one
step search}~) is used. In an earlier paper we had presented a detection
strategy, called
a {\em two step search}, that 
utilizes a hierarchy of template banks. It was shown that 
in the simple case of a family of Newtonian signals, an on-line 
 two step search was $\simeq 8$ times faster than an on-line one step 
search~(for initial LIGO).
 In this paper we extend the two step search 
to the more realistic case of zero spin ${\rm post}^{1.5}$-Newtonian wave 
forms. We also present formulas
for detection and false alarm probabilities which take statistical
correlations into account. We find that for the case of a
 ${\rm post}^{1.5}$-Newtonian family of
templates and signals, an on-line two step search  
 requires $\sim 1/21$ the computing power that would be required for the
 corresponding on-line one step search. This reduction is achieved when 
signals having  strength $S = 10.34$ are required to be detected with 
a probability
 of 0.95, at an average of one false event per year, and the noise power
 spectral density used is that of advanced LIGO. For initial
LIGO, the reduction achieved in computing power is $\sim 1/27$
 for $S = 
9.98$ and the same probabilities for detection and false alarm as
 above.
\end{abstract}

\section{Introduction}
\label{intro}
A radiation reaction driven inspiral of  a binary composed of 
compact massive objects~(Neutron Stars, Black Holes) would emit gravitational
waves that would lie, during the last few minutes before merger, in the sensitive bandwidth of
Laser interferometric detectors like LIGO~\cite{LIGO}, VIRGO~\cite{VIRGO}
and GEO600. Even though most such events will produce a signal amplitude 
well below the noise rms at any given instant, the predictability of such
wave forms would allow the use of pattern matching techniques
like matched filtering to  considerably improve their chances of
 detection~\cite{Thorne87}. In
matched filtering, the detector output is passed through a filter that is 
matched to the expected signal wave form in some optimal sense. If the 
maximum of the output crosses a pre-determined threshold, a signal is 
declared to be present in the data with a time of arrival given by the 
location of the maximum. The filtering of the detector output
 can, of course, be substituted with a cross-correlation against a 
suitable template wave form that is matched to the expected signal.

Generically, the wave form from an inspiraling binary is an amplitude and 
phase modulated sinusoid both of whose instantaneous frequency 
and amplitude increase as the two bodies proceed towards merger. The 
signal becomes ``visible'' in the output of a detector when its 
instantaneous frequency exceeds the lower frequency cutoff of the output's
 bandwidth. This
moment can be taken as the time of arrival of the signal at the detector.
Such a cutoff is required, for instance, in the case of ground
 based detectors because of excessive  seismic noise  at low frequencies. 
There would, of course, be an unknown phase offset at the time of arrival.
In addition, the signal would be characterized by the masses
and spins of the two components, the distance to the binary and
 geometrical factors like the direction to the binary and the orientation 
of the orbital plane.

Thus, if matched filtering is used for
such signals, it would be necessary to employ a bank of filters~(or template wave forms) corresponding to different values of the signal parameters
 mentioned above. 
One would then compare the maximum over all the filtered
 outputs with a threshold.
This strategy is usually called a {\em one-step search}.
Even though the time of arrival, initial phase and distance can be handled  
easily, a one-step search would still be a computationally expensive
 proposition. Estimates~\cite{bowen96,apost96} of the computational power 
that is required for an on-line one-step search using post-Newtonian templates
turn out to be 
 $\sim 200$~Gflops~(Giga flops) or higher.
This is with the omission of various other
signal processing overheads which can be expected to be present
 in a realistic situation. Therefore, it is
  desirable to have computationally less expensive 
detection strategies without, however, compromising too much
 on the performance afforded by matched filtering. One such strategy, called a {\em two step
hierarchical search}, was investigated by us in an earlier work~\cite{MD96}.
We found that this strategy reduces the computational cost of detection
significantly {\em without} losing out on the performance of a 
one step search. This is because a two-step search utilizes information
that was present in a one-step search but which was neglected. Namely, the
correlation between templates which allows a coarse scan of the parameter
space to predict the location of a threshold crossing peak among the filtered
outputs.
 
Our analysis was restricted in~\cite{MD96}~(henceforth referred to as MD96)
to the case of Newtonian wave forms and the noise
power spectral density used was that of the initial LIGO. The main result of
our analysis was that a two step hierarchical search is $\sim 8$ times faster
than the corresponding one step search. This gain was achieved when the 
detection probability desired was 0.95 for a signal to noise ratio of
 $ 8.8\sigma$ at an
 average rate of false events of 1/y.
 The Newtonian template family
will, however, not be good enough for the detection of the true signal 
wave form~\cite{apost95}. It was chosen in MD96 in order to keep the 
analysis simple since
 that work was in the nature of a  first estimate and several other issues 
needed to be highlighted. This paper is an extension of the two step search
to a more realistic family of signals and templates, namely, the 
${\rm post}^{1.5}$-Newtonian family.  

Our choice  is motivated by the result of 
Apostolatos~\cite{apost96} that a ${\rm post}^{1.5}$-Newtonian template family, 
having spin parameters
$\beta = \sigma = 0$, is adequate for the detection of signals up to~(and
 possibly beyond) 
${\rm post}^{2}$-Newtonian order, even for maximally spinning systems. However,
 this holds for spins that are aligned with the orbital 
angular momentum. In general, the signal from a misaligned system would 
suffer significant phase and amplitude modulations that can considerably
reduce the SNR, for some detector-binary geometries, if non-spinning
templates are used.
The larger the opening angle between the orbital and total angular momentum,
the larger is the fraction of detector-binary geometries which would be
 lost. But for moderate opening angles~($\sim 25^\circ$ or less),
 a sizable fraction of signals can still be detected with the non-spinning 
templates. Therefore, a non-spinning ${\rm post}^{1.5}$-Newtonian 
template family appears to be a realistic one to use.
 We choose our family of 
signals also to be the same since, as mentioned above, even higher order signal
wave forms may be detectable using this family of templates. This choice of
templates and signals should provide a realistic model for the 
assessment of a two step hierarchical search while keeping the analysis
relatively simple. In the following we will refer to non-spinning 
${\rm post}^{1.5}$-Newtonian wave forms as simply ${\rm post}^{1.5}$-Newtonian
ones.

The main result of this paper is summarized in Tables~\ref{initligotable}
 and~\ref{advligotable}. In Table~\ref{initligotable}, the noise power 
spectral density~(p.s.d.) expected for the initial
 LIGO~\cite{FC93} has been used
while in Table~\ref{advligotable}, the p.s.d.~used is that which is expected for
the advanced LIGO~\cite{CUTFLAN}. 
Columns seven and eight of each table show the computational power 
required for an on-line two-step search~($C_{\rm online}^{(2)}$)
 and an on-line one-step~($C_{\rm online}^{(1)}$) respectively
for the {\em same} performance parameters. That is, a 
detection probability of 0.95 for all signals having a {\em strength} 
$S_{\rm min}$~(as given in the captions of the tables) and an average
false alarm rate of 1~false~event/y. These values of the computational
requirement have been obtained for various lengths of the input data segment
which are tabulated in the first column. $\xi_{\rm max}$ is the length of
the template having the lowest values for the binary masses, $m_1$ and
$m_2$, which we choose to be $m_1 = m_2 = 0.5\; M_\odot$. The highest masses
that we have used in our analysis are £$m_1 = m_2 = 30.0\;M_\odot$.
 These results show that
a two step hierarchical search can reduce computational requirements by
 about a factor of $\sim 25$ in a realistic scenario. 

In MD96, the formulae used for detection and false alarm probabilities used
the assumption of statistical independence between certain random variables.
This assumption fails when templates are placed very closely and we were, thus,
limited from exploring the small spacing case more thoroughly. In the 
present paper, we present a much improved formula for the detection probability
that reproduces the Monte Carlo results quite well. It also suggests an 
alternative approach to Monte Carlo simulations for parameter estimation which
should be orders of magnitude faster than the conventional approach but 
further investigations in this direction are postponed to a later work. We 
also show, somewhat qualitatively, that the assumption of statistical 
independence is justified as far as the false alarm is concerned. Thus, the
results that we have obtained
 can be considered to be fairly accurate within the
approximations that have been made due to other reasons, like the non-trivial
boundary of the space of interest and the location dependence of the 
intrinsic ambiguity function.

The rest of the paper is organised as follows. In Section~\ref{MLD} we apply
the method of {\em maximum likelihood detection} to the  
${\rm post}^{1.5}$-Newtonian family of signals. This is the rigorous 
formalism behind the matched filtering algorithm mentioned above. We end 
this section with the derivation of a {\em test statistic}
 whose value determines the choice between detection and non-detection.
In Section~\ref{dist}, we
study the probability distribution functions of the test statistic. Formulas
for the detection and false alarm probability are derived.
In Section~\ref{tplace1} the problem of optimaly placing templates in a one-step
search is investigated. We obtain some approximations regarding the 
placement geometry, number of templates etc.  
Section~\ref{twostp} is devoted to the two-step hierarchical search
and also contains the results of this paper.
 We conclude with Section~\ref{conclude}.

\section{ Maximum Likelihood detection of ${\bf post}^{1.5}$-Newtonian
 signals  }
\label{MLD}

The method of {\em maximum likelihood} detection entails the maximization
of the {\em posterior} probability $p(x(t)\, | \, \Theta)$ over the signal
parameters $\Theta$. Here, $x(t)$ is a segment of the output of a detector and 
$p(A\, |\, B)$ denotes the conditional probability of $A$ given $B$.
Actually, one should maximize $p(\Theta \, | \,x(t))$ because it is $x(t)$  which is given but when 
our {\em a priori} knowledge of the frequencies with which various values of
 the parameters can occur is negligible,
 this is  equivalent to the maximization of $p(x(t)\, |\, \Theta)$ .
The maximum is then compared with a fixed threshold and a detection is 
announced if the threshold is crossed. Maximum likelihood detection also achieves,
the highest {\em average} detection probability~(averaged over $\Theta$)
 for a given false alarm probability. Actually, this statement is only true
approximately but the approximation becomes better as the signal to noise 
ratio becomes larger~\cite{Helstrom}.

If the detector noise is assumed to be a stationary Gaussian process, defined
by a~(one-sided) power spectral density $S_n(f)$, the above maximization 
reduces to the maximization, over $\Theta$, of the quantity
$$
\int_{-\infty}^{\infty}{d f\over S_n(f)}\, 
 \widetilde{x}(f)\widetilde{s}^\ast(f;\, \Theta)  - 
{1\over 2} \int_{-\infty}^{\infty}{d f\over S_n(f)}\, 
 \widetilde{s}(f;\, \Theta)\widetilde{s}^\ast(f;\, \Theta)\;,
$$
 where  a `$\sim$' stands for the Fourier transform of the corresponding
time domain function and $s(t;\, \Theta)$ stands for a member of the signal
family.
 This motivates the definition of an inner product,
\begin{equation}
\langle \, u(t),\, v(t)\, \rangle =   
\int_{-\infty}^{\infty}{d f\over S_n(f)}\, 
 \widetilde{u}(f)\widetilde{v}^\ast(f) \;,
\end{equation}
and a corresponding norm,
\begin{equation}
\| u\| = [ \langle \, u,\, u\, \rangle ]^{1/2} \; .
\end{equation}
Thus, maximum likelihood detection involves the computation of,
\begin{equation}
\Lambda = \max_{\Theta}\left[ \langle \, x(t),\, s(t;\, \Theta)\, 
\rangle - {1\over 2}
\langle \, s(t;\, \Theta),\, s(t;\, \Theta)\, \rangle \right] \;.
\end{equation}
This quantity is known as the {\em test statistic}. The test statistic is then
compared with a pre-determined threshold to decide whether the given $x(t)$
contains a signal or not. It should be noted that the properties of 
stationarity and Gaussianity are only approximations for the noise that will
be present in the interferometric detectors. However, these approximations
are expected to be quite good and we will assume such a noise in the following.

Some clarification should be made here regarding the meaning
of a detector output in the context of interferometric gravitational wave
detectors. The final output at the photodetector would contain the
response of the detector~(as calculated using the geodesic deviation
equation) convolved with the detector's transfer function~\cite{KLM}~(which depends
on the way the detector is configured, i.e., the kind of recycling used, the
reflectivities of mirrors etc.). This signal would be buried in noise that 
would  be {\em white} Gaussian noise if photon shot noise were the 
only source of noise. So, strictly speaking, the detection
strategy should be for the detection of this {\em convolved} signal in white
 noise. 
However, it is easy to prove that this is equivalent to the detection of the
deconvolved signal in a noise which has a power spectral density $S_n(f)$
such that when this combination is ``passed'' through a {\em noise free}
 interferometer, the output is the convolved signal and {\em white}
 noise, as would be observed actually. Clearly, $S_n(f)$ should be 
inversely proportional
to the modulus squared of the detector's transfer function. When the noise
at the output is not white, as would be the case in practice, $S_n(f)$ would be
 the power spectral density of the actual noise
 divided by the modulus squared of the detector's transfer function.
Henceforth, by the detector output we would mean the {\em deconvolved} output
which would have noise with a power spectral density given by $S_n(f)$ and
the signal would be just the bare response of the interferometer i.e.,
 the relative strain produced in the two arms as computed using the geodesic
deviation equation. 
 
\subsection{ The ${\rm \bf post}^{\bf 1.5}$-Newtonian signal }
\label{pnsignal}

In the case of coalescing binary signals expressed
up to the ${\rm post}^{1.5}$-Newtonian order~(spin parameter $\sigma = 0$),
the signal parameters are,
 $\Theta = ({\cal A}, \; t_a, \; \Phi, \; \tau_0, \; \tau_{1.5})$.
The parameter $\cal A$ is the~(nearly) constant part of the amplitude of the
 signal that takes into account  the distance to the binary and the relative
orientation of the detector and the $TT$ coordinate frames. The rapid rise of 
 power in the seismic noise towards lower frequencies would require that the 
detector output be band pass filtered with a cutoff at some low frequency~(usually assumed to be 40~Hz
for the initial LIGO and 10~Hz for the advanced LIGO). Similarly,
a cutoff $f_c$
 will also be required at the high frequency end because of a rise in
photon shot noise. Usually, $f_c$ is taken as $f_c = 1000$~Hz.
Thus, loosely speaking,
the signal wave form from an inspiraling binary would ``start'' in the output 
of the detector at the time when its instantaneous frequency crosses $f_a$. 
This instant is called the time of arrival of the signal and is denoted by
 $t_a$. The phase of the wave form at $t= t_a$ is denoted by $\Phi$. The 
remaining parameters have the dimension of time and depend on the masses of 
the binary components. They have been called as {\em intrinsic} parameters
of the wave form in contrast to $t_a$ and $\Phi$ which are known as 
{\em extrinsic } parameters.

Actually, the ${\rm post}^{1.5}$-Newtonian wave form should be characterized 
by {\em three} intrinsic parameters, $(\tau_0,\, \tau_1,\, \tau_{1.5})$.
The subscripts of $\tau$ denote the post-Newtonian order at which 
that parameter occurs.
 Thus, $\tau_0$ is the Newtonian chirp time characterizing the 
lowest order wave form obtained using the quadrupole formalism and 
the ${\rm post}^{1}$-Newtonian wave form would have $\tau_0$ and $\tau_1$
as its intrinsic parameters. However,
 we have assumed the spins of the binary components to be zero and, hence, 
we are left with only two {\em independent} intrinsic parameters which we
have chosen as $\tau_{1.5}$ and $\tau_0$ because 
they can be easily inverted to obtain the masses. In the expression for
the wave form, however, we retain $\tau_1$ with the understanding that it
is dependent on $\tau_{1.5}$ and $\tau_0$.
 In an {\em approximate} sense, $\tau_0 + \tau_1 - \tau_{1.5}$
can be taken to be the time left to the
  final merger of the binary, starting from 
$t=t_a$. 

In the following we will deal with the {\em restricted} form of the ${\rm post}^{1.5}$-Newtonian signal in which post-Newtonian corrections are 
applied to only the phase of the signal. The restricted wave form at 
any post-Newtonian level is expected to be a good model for the correct 
wave form at the same level~\cite{CUTFLAN}. 
The restricted  ${\rm post}^{1.5}$-Newtonian signal is,
\begin{equation}
h(t;\, \Theta) = {\cal A}\, a(t -t_a;\, \tau_0)
\cos\left[ \int_{-\infty}^{t-t_a}\!\!\!\! d t^\prime\, f(t^\prime;\,
 \tau_0,\tau_{1.5}) + \Phi\right] \;, 
\label{thesignal}
\end{equation}
where,
\begin{equation}
a(t) = \left( 1 - {t\over\tau_0} \right)^{-1/4}\;,
\end{equation}
and $f(t;\,\tau_0,\tau_{1.5})$, the instantaneous frequency of the signal, is given by an implicit equation,
\begin{equation}
t = \tau_0+\tau_1-\tau_{1.5}-\left( \tau_0 x^{-8/3}+\tau_1 x^{-2}
-\tau_{1.5} x^{-5/3} \right)  \; , \label{tdwaveform}
\end{equation}
where $x = f(t;\,\tau_0,\tau_{1.5})/f_a$.
The chirp times are given by the following expressions~($G=c=1$),
\begin{eqnarray}
\tau_0 & = & {5\over 256}{\cal M}^{-5/3}(\pi f_a)^{-8/3} \;, \\ 
\tau_1 & = & {5\over 192\mu(\pi f_a)^2} \left({743\over 336}+{11\over 4}\eta
 \right)  \;,\\
\tau_{1.5} & = & {1\over 8\mu}\left( {M \over \pi^2 f_a^5} \right) \;,
\end{eqnarray}
where, $M$ is the total mass of the binary, $\mu$ is the reduced mass, $\eta =
 \mu/M$ and ${\cal M} = (\mu^3 M^2)^{1/5}$ is the {\em chirp mass}.
We have chosen $\tau_0$ and $\tau_{1.5}$ as our independent parameters. 
In terms of these parameters,
\begin{eqnarray}
\tau_1 & = &{3715\over 4032}\left({4\over 5 \pi^2}\right)^{1/3}
 \tau_0^{1/3}\tau_{1.5}^{2/3}+ {11\over 24 f_a} \tau_0 \tau_{1.5}^{-1} \;,\\ 
\mu & = & {1\over 16 f_a^2}\left( {5\over 4\pi^4 \tau_0\tau_{1.5}^2}
 \right)^{1/3} \label{mufromip}\;, \\
M & = & {5\over 32}{ \tau_{1.5}\over \pi^2 \tau_0 f_a} \label{Mfromip} \; .
\end{eqnarray}
 Note that if $\tau_1$ were used instead of
 $\tau_{1.5}$, then the inverse relations for $m_1$ and $m_2$ would be more
complicated and would have to be solved numerically.

 In Fig.~\ref{initligospcofint} and
Fig.~\ref{advligospcofint}, we have shown the
$(m_1,\, m_2)$ plane~(the binary masses), $0.5 M_\odot \leq m_1, m_2 \leq 
30 M_\odot$, mapped into the $(\, \tau_{1.5},\, \tau_0\, )$ space for 
$f_a = 40$~Hz~(initial LIGO) and $f_a = 10$~Hz~(advanced LIGO) respectively.
The boundaries of the region, which we call the {\em space of interest} 
following Apostolatos~\cite{apost96}, can be obtained easily. The equation for
the curve $AC$, corresponding to $m_1 = m_2$ or $M = 4\mu$, can be found by 
directly substituting for $\mu$ and $M$ from~(\ref{mufromip}) and Eq.~(\ref{Mfromip}).
To obtain the other two segments, $AB$ and $CB$, we use the following
 expression,
\begin{equation}
{256\over 5} \tau_0 (\pi f_a)^{8/3} = {(m_1 + m_2)^{1/3}\over m_1 m_2} \; .
\end{equation}  
For every  value of $\tau_0$, the line of constant $\tau_0$ will intersect
$AB$ at a point where one of the masses, say $m_2$, is $0.5 M_\odot$. Similary,
for $BC$, one of the masses, say $m_1$, would be  $30.0 M_\odot$
at the point of intersection. The point where $AB$ and $BC$ meet falls on the
$\tau_0 = \tau_0(30.0\, M_\odot,\, 0.5\, M_\odot\, )$ line. Thus, for values
of $\tau_0$ larger than this, one of the masses in the above equation can be
set to $0.5 M_\odot$ and the equation can be solved for the other mass to 
yield the value for $\tau_{1.5}$ at the point of intersection.
 For smaller values of $\tau_0$, one of the
masses should be set to $30 M_\odot$ and $\tau_{1.5}$ be obtained as before.
Thus, given a value of $\tau_0$, the two limits of $\tau_{1.5}$ can be 
computed. This allows the area of the space of interest $A$
 to be computed using
a standard 2D Quadrature algorithm~(we use D01DAF of the NAg library). The
area of the space of interest thus obtained is,
\begin{equation}
A = \left\{ \begin{array} {lr}
50.174~{\rm sec}^2 & \mbox{for initial LIGO}\\
20389.542~{\rm sec}^2 & \mbox{for advanced LIGO}
\end{array} \right.\;.
\label{areaval}
\end{equation}

We can write the R.H.S. of Eq.~(\ref{thesignal}) as
\begin{eqnarray}
h(t;\, \Theta) & = & {\cal A} h_0(t-t_a;\,
 \tau_0,\tau_{1.5} ) \cos\Phi + {\cal A} h_{\pi/ 2}
(t-t_a;\, \tau_0,\tau_{1.5} ) \sin\Phi \;, \\
h_0 (t;\, \tau_0,\tau_{1.5} ) & = & a(t;\, \tau_0) \cos\left(
 \int_{-\infty}^{t}\!\!\!\! d t^\prime\, f(t^\prime;\,
 \tau_0,\tau_{1.5}) \right) \; , \\
h_{\pi/ 2}(t;\, \tau_0,\tau_{1.5} ) & = & a(t;\, \tau_0) \cos\left(
 \int_{-\infty}^{t}\!\!\!\! d t^\prime\, f(t^\prime;\,
 \tau_0,\tau_{1.5}) + {\pi\over 2} \right) \;.
\end{eqnarray}
This representation will be useful in what follows.

The Fourier transform of $h(t;\,\Theta)$ can be computed 
using the stationary phase approximation. For our purpose, it suffices
to give its overall form here, the details being available in other 
sources~\cite{apost96,CUTFLAN}. For $f> 0$,
\begin{equation}
\widetilde{h}(f;\, \Theta )  \propto  {\cal A}
f^{-7/6} \exp[ -2\pi i ( f t_a +  \sum_i\tau_i \psi_i (f)) +i\Phi ]\; , \label{statphase}
\end{equation}
where the index $i \in\{ 0, 1, 1.5\}$. 
 The functions $\psi_i$ are independent of the signal parameters.
For $f < 0$, the transform is constructed using the Hermitian property
of the Fourier transform, $\widetilde{h}(f) = \widetilde{h}^\ast(-f)$, since
$h(t,\,\Theta)$ is a real function.

\subsection{ The test statistic and its computation }
\label{tsstat+comp}

Following the brief outline given earlier,
 the maximum likelihood detection strategy for 
${\rm post}^{1.5}$-Newtonian signals would consist of the computation of
a test statistic $\Lambda$ given by,
\begin{equation}
\Lambda = \max_{\Theta} \left[ 
 \langle\, x(t),\, h(t;\, \Theta) \rangle -{1/2}\langle\, 
h(t;\, \Theta),\, h(t;\, \Theta)\, \rangle \right]\;.
\label{originalts}
\end{equation}
For the sake of convenience in the following, we adopt the notation 
\begin{eqnarray}
\theta^\prime &= &(t_a, \tau_{1.5}, \tau_0)  \;, \\
\theta &= & (\tau_{1.5}, \tau_0) \;.
\end{eqnarray}
Occasionally, we will also break up $\theta^\prime$ as $t_a$ and $\theta$.
The maximization over the parameters ${\cal A}$ and $\Phi$ can be carried out
 analytically to yield,
\begin{equation}
\Lambda = \max_{\theta^\prime}
{ \langle\, x,\, q_0\, \rangle^2 + \langle\, x,\, q_{\pi/ 2}\, \rangle^2
 -\langle\, x,\, q_0\, \rangle
\langle\, x,\, q_{\pi/ 2}\, \rangle \langle\, q_0,\, q_{\pi/ 2}\,\rangle
 \over 1 - \langle\, q_0,\, q_{\pi/ 2} \rangle^2 } \; ,
\end{equation}
where, 
\begin{eqnarray}
q_0(t;\, \theta) & = & {\cal N}_0^{-1} h_0(t;\, \theta)
 \;, \\
q_{\pi/ 2}(t;\, \theta) & = & 
{\cal N}_{\pi/ 2}^{-1} h_{\pi/ 2}(t;\, \theta) \;, \\
{\cal N}_0 & = & \| h_0(t;\, \theta)\|\;, \label{norm0}\\
{\cal N}_{\pi/ 2} & = & \| h_{\pi/ 2}
(t;\, \theta)\| \;.\label{norm1}
\end{eqnarray}
${\cal N}_0$ and ${\cal N}_{\pi/ 2}$ are normalization constants that are
 chosen as above for later convenience. Since $t_a$ occurs in the phase of the
Fourier transforms of $h_0(t-t_a;\, \theta)$ and 
$h_{\pi/ 2}(t-t_a;\, \theta)$, ${\cal N}_0$ and
 ${\cal N}_{\pi/ 2}$ are independent of $t_a$. We call $q_0(t;\theta)$ and $q_{\pi/ 2}(t;\theta)$ collectively as the {\em template} located at
 $\theta$ and $q_0$, $q_{\pi/ 2}$ themselves as the ``quadrature''
components of the  template.

It can be shown, using the stationary phase Fourier 
transform given in Eq.~(\ref{statphase}), that
\begin{eqnarray}
 {\cal N}_0 & = & {\cal N}_{\pi/ 2}\;, \label{equalnorms}\\
\langle\, q_0,\, q_{\pi/ 2}\, \rangle & = &0 \label{ortho} \;. 
\end{eqnarray}
In general, if
\begin{equation}
{\cal N}_\Phi = \| h(t;\,{\cal A} = 1, \Phi, \theta)\|\;,
\label{normphi}
\end{equation}
then it can be shown using Eq.~(\ref{statphase}) that 
\begin{equation}
{\cal N}_\Phi = {\cal N}_0 = {\cal N}_{\pi/ 2} \;.
\label{allnormsequal}
\end{equation} 
Note that ${\cal A} = 1$ in the above.
 Eqs.~(\ref{ortho}), (\ref{allnormsequal})
also hold to a very good approximation 
when the numerically computed Fourier transforms of the wave forms 
are used~(the typical variation in ${\cal N}_\Phi$ is $< 1\%$ 
over $\Phi = [0,\,2\,\pi]$).
By this we mean that first the wave forms are 
generated in the time domain using Eq.~(\ref{tdwaveform}) and then
the Fourier transforms are computed using an FFT. Henceforth, we take 
${\cal N}_\Phi$ to be independent of $\Phi$ and denote it simply by ${\cal N}$.
However, $\cal N$ does depend on the chirp times via the proportionality 
constant in Eq.~(\ref{statphase}).

 At this
point it is convenient to define a quantity which we call the {\em strength}
$ S $ of a signal~\cite{SD94,MD96},
\begin{equation}
S = {\cal A} {\cal N}\;.
\label{strengthdef}
\end{equation} 
Henceforth, we use $S$ instead of $\cal A$ to parametrise a signal. This
quantity is essentially the same as the signal to noise ratio $[S/N]$ as
defined in~\cite{CUTFLAN}.

As a consequence of Eq.~(\ref{equalnorms}) and Eq.~(\ref{ortho}),  the test statistic in Eq.~(\ref{originalts}) reduces to
\begin{eqnarray}
\Lambda & = &\max_{\theta^\prime} \left[ C_0^2(x;\, \theta^\prime) + C_{\pi/ 2}^2(x;\, \theta^\prime) \right]^{1/2}
\; \label{teststat},\\
& &C_0(x;\, \theta^\prime)  =  \langle\, x(t),\, q_0(t-t_a;\, \theta)\, \rangle  \; ,\\
& & C_{\pi/ 2}(x;\, \theta^\prime)  = \langle\, x(t),\, q_{\pi/ 2}
(t-t_a;\, \theta)\, \rangle \; .
\end{eqnarray}
The square root in Eq.~(\ref{teststat}) is, strictly speaking, not necessary but we retain it in order
to make our analysis conform to some of the existing literature. 

It has not
been possible so far to proceed further analytically with the
 maximization involved in Eq.~(\ref{teststat}). Although several numerical 
methods are available for the maximization of functions~\cite{NREC}, such
methods tend to fail
  when the function involved has many local maxima~(as would be the case
here), since  most of these methods have a tendency to converge on one 
of the local
 maxima rather than the required global maximum. 
 Also, given the very small
expected event rate, it is necessary to have a good {\em a priori} estimate
of the false alarm and detection probabilities and these quantities are not easy
to compute for such methods. The only alternative left is a search for the
maximum over a grid of points in parameter space. This method is also
simple enough that its performance can be analysed theoretically to a large
extent. The practical implementation of such a method,
 called a {\em one-step search},
 can be motivated as follows.

For a fixed $\theta = \theta_0$,
 the computation of
$C_0(x;\, t_a, \theta_0)$~(or $C_{\pi/ 2}(x;\, t_a, \theta_0)$), as a function of 
$t_a$, is equivalent to taking the linear correlation~\cite{BRCWELL}
 of $x(t)$ with the
template $q_0(t;\, \theta_0)$~(or $q_{\pi/ 2}$ in the case of
$C_{\pi/ 2}$). Since correlations can be computed efficiently using the
Fast Fourier Transform~(FFT)~\cite{Brigham}, 
the maximization over $t_a$ alone is 
straightforward: The detector output would be sampled at a rate greater than or
equal to the Nyquist rate~(in this case $\sim 2000$ Hz) yielding a discrete
time series $\overline{x} = (\, x_0, x_1, .\, .\, .\, x_{N-1}\, )$. This time
series can then be correlated, using FFTs, 
 with analogous time series $\overline{q}_0$ and $\overline{q}_{\pi/ 2}$
 for the two quadrature components. However, the whole output of such
a correlation cannot be used since the use of a FFT will yield
a {\em circular} correlation instead of a linear one. It is, therefore,
 necessary to
have a {\em padding} of zeroes at the end of each template time series. Let the
duration of this padded part be $T_P$~sec for some template.
 Then, for that template, only the first $T_P$~sec of
the correlation outputs will be the result of a linear correlation and the
rest would have to be discarded. Note that $T_P$ will depend on the template 
parameters since the duration of the wave form is parameter
 dependent~(approximately equal to $\tau_0+\tau_1-\tau_{1.5}$).
 Let the longest duration
among the template wave forms be $ \xi_{\rm max}$~sec~(not be confused with
the Newtonian chirp time used in MD96). Then, the shortest linear
correlation will have a duration of $T_P^0 =T-\xi_{\rm max}$, 
where $T$ is the duration of the of the input time series $\overline{x}$.
 We will assume in this
paper that only the first $T_P^0$~sec will be kept in each correlation output
even if the duration of a template wave form is much less than $ \xi_{\rm max}$.
This appears as a wasteful procedure and a better use of computational 
resources may be possible. However, we do not investigate this issue here since
it is not directly relevant to this paper.

Having obtained the correlations with $\overline{q}_0$ and 
$\overline{q}_{\pi /2}$, 
the first $T_P^0$~sec of each correlation should be squared, 
corresponding samples of the two outputs should be added and the 
square root taken  to yield a single
time series. We call this final time series obtained for some template 
parameters $\theta$ as 
the {\em rectified output}~\cite{MD96} of the template $\theta$. 
Such rectified outputs can be
 constructed for several points in the~($\tau_{1.5}$, $\tau_0$) space and
the maximum found over each of them. Finally, the maximum of all these maxima
will yield an approximation to the test statistic $\Lambda$. This, essentially,
is the scheme of a one step search. We call the set of points 
in~($\tau_{1.5}$, $\tau_0$) space for which rectified outputs are generated as 
the {\em bank of templates}. The coordinates of any rectified output 
sample is given by $\theta^\prime$ while that of a template is given by
 $\theta$.

\section{ Distributions of the test statistic}
\label{dist}

To quantify the performance of the detection strategy described above, we
need to calculate the probability of a false alarm as well as that of 
detection for a given signal. The former is defined as the probability of
the test statistic crossing the threshold when a signal is absent in $x(t)$.
The latter is the probability of the test statistic crossing the threshold
when a signal is present. It should be noted that the detection probability 
need not be the same for all signals in a signal family. 
For instance, large amplitude signals should,
obviously, have a larger detection probability than weaker ones. We
denote the detection probability of a signal with parameters $\Theta$ by
$Q_d(\eta;\, \Theta)$ where $\eta$ denotes the threshold. We denote the
false alarm probability by $Q_0(\eta)$. If the probability density function
of $\Lambda$, the test statistic, be $p_0(\Lambda)$ in the absence of a signal
and $p_1(\Lambda;\, \Theta)$ in the presence of a signal, then,
\begin{eqnarray}
Q_0(\eta) & = & \int_\eta^\infty p_0(\Lambda)\, d\Lambda \; , \\
Q_d(\eta;\, \Theta) & = & \int_\eta^\infty p_1(\Lambda;\, \Theta)\, 
d\Lambda \; .
\end{eqnarray}
In order to construct $p_0$ and $p_1$, we start at the lowest level, namely,
the density functions of a single sample of a rectified output.

Let $z(\theta^\prime)$ be a sample of some rectified output. Now,
this sample will be of the form, $z(\theta^\prime) = [\,C_0^2(\theta^\prime) + 
C_{\pi/ 2}^2(\theta^\prime)\,]^{1/2}$, where the dependence on $x(t)$ has not been
shown explicitly. As mentioned before, both $C_0(\theta^\prime)$ and 
$C_{\pi/ 2}(\theta^\prime)$
are obtained from correlations involving the detector output. Thus, they are 
linear combinations of the samples of $x(t)$ and it follows that their 
marginal probability density function is a Gaussian since the noise is 
assumed to be a Gaussian random process. In the presence of a signal 
$h(t;\, S, \theta^\prime_s, \Phi_s)$,
their mean values would be,
\begin{eqnarray}
\overline{C}_0(\theta^\prime) & = &  \langle\,
h(t;\, S,\theta^\prime_s , \Phi_s, ), \,
q_0(t-t_a;\, \theta)\, 
 \rangle\;, \label{meanc0}\\
\overline{C}_{\pi/ 2}(\theta^\prime) & = & 
 \langle\, h(t;\, S, \theta^\prime_s, \Phi_s), \,q_{\pi/ 2}(t-t_a;\, \theta)\, 
 \rangle\;, \label{meanc1}
\end{eqnarray}
while in the absence of a signal, they will have zero means. 
 With our choice of 
${\cal N}_0$ and ${\cal N}_{\pi/ 2}$ in Eqs.~(\ref{norm0}),~(\ref{norm1}), it
can be shown that the variance of $C_0(\theta^\prime)$ and 
$C_{\pi/ 2}(\theta^\prime)$ is unity. It can
also be shown, using Eq.~(\ref{ortho}), that $C_0(\theta^\prime)$ and 
$C_{\pi/ 2}(\theta^\prime)$ are statistically independent of
each other. Note that, in general, the covariance of
$C_0(\theta^\prime_1)$ and
$ C_{\pi/ 2}(\theta^\prime_2)$ need not vanish for
$\theta^\prime_1 \neq \theta^\prime_2$.
 Given these properties for $C_0$ and $C_{\pi/ 2}$, it 
follows that the marginal  probability density function of a rectified output
sample is a Rician $Ri(z)$ when a signal is present and a Rayleigh $R(z)$
in the absence of a signal,
\begin{eqnarray}
Ri(z) & = & z \, \exp\left[-{1\over 2}(z^2+d^2)\right] I_0(z d) \; , \\
R(z) & = & z\, \exp\left[{z^2\over 2}\right] \; ,\label{rayleigh}
\end{eqnarray}
where, 
\begin{equation}
d^2 = d^2(\theta^\prime)  = \overline{C}_0^2(\theta^\prime) +
 \overline{C}_{\pi/ 2}^2(\theta^\prime)\;,
\label{d}  
\end{equation} and $I_0(x)$ is the modified
Bessel function of the first kind~(of order zero). 
For $z d \gg 1$, the asymptotic form of $Ri(z)$ is,
\begin{equation}
Ri(z) \sim  \sqrt{z\over 2 \pi d}\; \exp\left[-{1\over 2}(z-d)^2\right]\; .
\label{gaussapp}
\end{equation}
Thus, for $z\simeq d$, a Rician density goes over into a Gaussian.

Under the stationary phase approximation of Eq.~(\ref{statphase}), it can
be shown that $d(\theta^\prime)$ is independent of the phase $\Phi$ of the signal. 
Again, this is a good approximation for the exact numerical case.
We assume henceforth that $d(\theta^\prime)$ is independent of the signal 
phase $\Phi$.
Let
\begin{eqnarray}
H(\theta^\prime_p,\,\theta^\prime_q) & = &\left[\, \langle\, q_0(t-t_a^p;\,\theta_p ),\, 
q_0(t-t_a^q;\, \theta_q)\, \rangle^2 \right.\nonumber \\
& & + \left. \langle\,q_0(t-t_a^p;\, \theta_p)\, ,
 q_{\pi/ 2}(t-t_a^q;\,\theta_q )\, 
\rangle^2\, \right]^{1/2} \;.
\end{eqnarray}
Since $d(\theta^\prime)$ is 
almost independent of the signal $\Phi$, it follows from 
 Eqs.~(\ref{meanc0}),(\ref{meanc1}) and~(\ref{d}) that
\begin{equation}
d(\theta^\prime)  = S\, H(\theta^\prime,\,\theta^\prime_s) \;. \label{dforS}
\end{equation}
The  quantity $H(\theta^\prime_p,\, \theta^\prime_q)$ 
is related to the {\em ambiguity} function~\cite{Helstrom}. Note that since
$t_a$ occurs in the phase of a Fourier transform, $H(\theta^\prime_p,\,\theta^\prime_q)$
depends on $t_{a}^p$ and $t_{a}^q$ only through 
$\Delta t_a = t_{a}^p - t_{a}^q$.
 
In order to obtain the distribution of the test statistic $\Lambda$ we need
 to know the joint distribution of, in general, all the samples.
In the presence of a sufficiently strong signal,  however,
it can be expected that $\Lambda$ will occur almost always
only among those samples of the rectified outputs which have a high value of 
$d$. For a typical number of samples in a single rectified output of 
$\sim 10^5$, for instance, the location of $\Lambda$ is restricted 
significantly if some samples have $d \geq 5.0$. Otherwise, $\Lambda$ occurs
almost randomly anywhere within the output time series. Thus, to obtain the 
distribution of $\Lambda$ in the presence of a signal, we need to consider only
a restricted subset of all the samples, namely, those for which $d \gg 1$.
 The distribution of $\Lambda$ would
then be that of the maximum over this subset. We describe a general scheme
for choosing this subset below  but for the present, we assume that it
is given. As shown above~(see Eq.~(\ref{gaussapp})),
 each sample of this set would have a 
marginal distribution that is approximately a Gaussian. It is plausible that
the joint distribution of such samples can also be approximated by a 
multivariate Gaussian. This possibility can be investigated by computing the 
moments of the joint distribution and comparing them with those of a 
multivariate Gaussian. 

We can express a Rician
variable $Z = \sqrt{ X_1^2 + X_2^2}$, where $X_i$ is a Gaussian random 
variable with mean $\mu_i$, as
\begin{eqnarray}
Z & = & \left[ (\mu_{1} + \delta X_{1})^2 + (\mu_{2} + \delta X_{2})^2 \right]^{1/2}\;,\nonumber \\
& = & \sqrt{\mu_1^2 + \mu_2^2} \left[ 1 + {2(\mu_1\delta X_1 + \mu_2 \delta X_2)
\over \mu_1^2 + \mu_2^2} + {\delta  X_1^2 + \delta X_2^2 \over 
\mu_1^2 + \mu_2^2} \right]^{1/2}\; ,\label{expanded}
\end{eqnarray}
where $\delta X_{1}$ and $\delta X_{2}$ are zero mean Gaussian random variables
with unit variances. In the above expression and in the following, we will
follow the customary practice of denoting a random variable by a capital letter
while denoting its value in a particular realization by the corresponding
smaller case letter.
The probability that $ 2(\mu_1\delta X_1 + \mu_2 \delta X_2) + \delta  X_1^2 + \delta X_2^2 $  be larger than  $\mu_1^2 + \mu_2^2$ can be 
obtained easily,
\begin{equation}
\hspace{-1.5cm} {\rm Pr} \left[ {2(\mu_1\delta X_1 + \mu_2 \delta X_2) +
\delta  X_1^2 + \delta X_2^2 \over 
\mu_1^2 + \mu_2^2} \geq 1 \right] = {1\over 2 \pi}
\int_0^{2 \pi} d\theta \exp\left[-{1\over 2}
d^2 ( \sqrt{ 1 + \cos^2[\theta] } - \cos[\theta] )^2 \right]\;,
\end{equation}
where, $d^2 = \mu_1^2 + \mu_2^2$. In our analysis, $d \sim 7.0$ or larger. 
For $d = 7.0$, the above probability 
is 0.008 and it decreases rapidly for higher values. Thus, only a small fraction realizations would be such that their binomial expansion in terms of 
$(2(\mu_1\delta X_1 + \mu_2 \delta X_2) +
\delta  X_1^2 + \delta X_2^2 )/(\mu_1^2 + \mu_2^2)$ would be non-convergent. 
It would be a good approximation to neglect such realizations and calculate
the moments of $Z$ by expanding the RHS of 
Eq.~(\ref{expanded}) in a binomial expansion and taking the ensemble average
for each term. 

The same argument goes through for multivariate moments also except for the 
fact that the
fraction of realizations for which all the components of the moment can be
expanded binomially will decrease as the number of different variates 
increases. For instance, if the third moment $\overline{ (Z_1 - a_1) (Z_2 -a2)
(Z_3 - a_3)}$ is required around some point $(a_1, a_2, a_3)$ then the 
fraction of cases for which the above expansion will be invalid, taking
$d \sim 7$ for all of them,  would be 
$\sim 0.008 \times 3$. In practice there would 
be a significant overlap of various cases since the $Z_i$ would be 
statistically correlated and this fraction would actually be less. In any case,
for low
order moments, this method would still furnish a good approximation since the
fraction  of realizations with non-convergent expansions is still small.
For higher values of $d$, the fraction of invalid expansions would go down 
rapidly bringing higher moments also under the purview of this method.

Now
consider the restricted subset of rectified output samples mentioned
above. Let this set be $\{ Z_1,\ldots,Z_m\}$ and the mean values of 
the Gaussian components, $\{(X_{11},X_{12}),\ldots ,(X_{m1},X_{m2})\}$,
 associated with these samples be $\{(\mu_{11},\mu_{12}),\ldots,(\mu_{m1},\mu_{m2})\}$. That is,
$Z_i = [ X_{i1}^2 + X_{i2}^2]^{1/2}$ and $\overline{X}_{i1} = \mu_{i1}$ and
$\overline{X}_{i2} = \mu_{i2}$. Let the strength of the signal be $S$.
Following the argument given above, we
can express a moment~(about mean values) of the joint distribution of $\{Z_i,
\ldots , Z_m\}$ as,
\begin{eqnarray}
{\rm E}\left[ (Z_1-d_1)^p\ldots (Z_m-d_m)^s\right] & =  & {\rm E} \left\{
\left[{\mu_{11}\delta x_{11} + \mu_{12}\delta x_{12} \over d_1}\right]^p \ldots 
\left[{\mu_{m1}\delta x_{m1} + \mu_{m2}\delta x_{m2}
 \over d_m}\right]^s\right\}\nonumber \\
& & + {\rm E}\left\{ O\left[ \delta x^{2(p+\ldots+s)}/ S\right]\right\}\;,
\label{moments}
\end{eqnarray}
where ${\rm E}[\;\; ]$ denotes an ensemble average. Note that the first term
is independent of $S$. Also, since $\delta X_{i1}$ and $\delta X_{i2}$ are Gaussian random variables, this term is a moment of a multivariate Gaussian
 distribution. The remaining terms  are inversely dependent on $S$ as is shown schematically above. In general, the lowest order
 correction to the Gaussian moment will have an $S^{-2}$ 
dependence for even moments and an $S^{-1}$ dependence for odd moments. 
 
Thus, for a sufficiently high $S$, the moments of the
joint distribution function of 
$\{Z_i,\ldots , Z_m\}$ are approximately the same as the moments of a 
multivariate Gaussian distribution. This implies that for $S \gg 1$, the joint distribution of the set $\{ Z_1,\ldots,Z_m\}$ is 
given by an $m$-variate Gaussian distribution. It is not easy to see
 how small the corrections to the Gaussian parts of the moments 
 should be for a given error in the detection probability. However, the
above argument provides a sufficiently strong motivation to proceed with
the multivariate Gaussian approximation to the joint distribution of 
 $\{ Z_1,\ldots,Z_m\}$. 
 In order to construct this
distribution, we need only compute the covariance matrix for 
$\{ Z_1,\ldots,Z_m\}$.

Suppose we have two sample $Z_i$ and $Z_j$ having coordinates 
$\theta^\prime_i = (t_a^i, \tau_{1.5}^i, \tau_0^i)$ 
and $\theta^\prime_j = (t_a^j, \tau_{1.5}^j, \tau_0^j)$
respectively. It is easy to show, using the stationary phase Fourier
transform, that the covariance matrix of $\{\, X_{i1},\, X_{i2},\, X_{j1},\, X_{j2}\,\}$
is, with the columns of the matrix in the same order,
\begin{equation}
{\bf C} = \left( {\begin{tabular}{llll}
1& $0$ & $r$ & $s$\\
0 & 1 & $ -s $& $r$ \\
$r$& $-s$ & 1 & 0 \\
$s$ & $r$ & 0 & 1 \\
\end{tabular}}\right) \; ,
\label{covmat}
\end{equation}
where, 
\begin{eqnarray}
r & = & \langle\, q_0(t-t_a^i;\, \theta_i),\,
 q_0(t-t_a^j;\,  \theta_j)\, \rangle \;, \label{r}\\
s & = & \langle\, q_0(t-t_a^i;\, \theta_i),\,
 q_{\pi/ 2}(t-t_a^j;\,  \theta_j)\, \rangle\label{s} \;.
\end{eqnarray}
Both $|r|$ and $|s|$ are less than unity. 
The above form of $\bf C$ is also approximately true when the numerically 
computed Fourier transforms of the templates are used. The covariance of 
$Z_i$ and $Z_j$ can now be computed using Eq.~(\ref{moments}) with the RHS 
truncated to the first term. We get,
\begin{equation}
\sigma_{ij} =
 {1\over d_i d_j}\left[r(\mu_{i1}\mu_{j1}+\mu_{i2}\mu_{j2})
+ s(\mu_{i1}\mu_{j2}-\mu_{i2}\mu_{j1})\right]
 \;.\label{covric}
\end{equation}
The same kind of calculation also yields, $\sigma_i = \sigma_j =
1$.  The covariance matrix, for the set $\{ Z_1,\ldots,Z_m\}$ above, can now
be computed using Eq.~(\ref{covric}). We can also express the 
covariance as 
 $\sigma_{ij} = \sqrt{r^2 + s^2}\, \chi$, where 
\begin{equation}
\chi = \tan^{-1}\left[\frac{\mu_1}{\mu_2}\right]+
\tan^{-1}\left[\frac{\nu_2}{\nu_1}\right]-
\tan^{-1}\left[\frac{s}{r}\right]\;.
\end{equation} 
Note that $\sqrt{r^2 + s^2}$ is
 precisely the quantity $H(\theta^\prime_i, \theta^\prime_j)$.

We list here the general expressions for the first three multivariate moments
 obtained using Eq.~(\ref{moments}), up to the 
lowest order correction to the Gaussian part.
 The algebra involved is tedious but a lot of it 
was automated using the symbolic computation package {\sc mathematica}. 
Let the three
rectified output samples be
 $Z_m = [X_1^2 + X_2^2]^{1/2}$, $Z_n = [Y_1^2 + Y_2^2]^{1/2}$ 
and $Z_o = [W_1^2 + W_2^2]^{1/2}$ and the covariance matrix for
$(X_1,X_2,Y_1,Y_2,W_1,W_2)$ be~(columns ordered in the same way),
\begin{equation}
{\bf C} = \left( {\begin{array}{llllll}
1& 0 & r_1 & s_1 & r_2 & s_2\\
0 & 1 &  -s_1 & r_1 &-s_2 & r_2 \\
r_1& -s_1 & 1 & 0 & t & u \\
s_1 & r_1 & 0 & 1 & -u & t \\
r_2 &  -s_2 &  t &  -u &  1 &  0 \\
s_2 &  r_2 &  u &  t &  0 &  1
\end{array}}\right) 
\end{equation}
The moments are,  
\begin{eqnarray}
\overline{(Z_m-d_m)} & = & \frac{1}{2 d_m}\;, \\
 \overline{(Z_m-d_m)^2} & = & 1 - \frac{1}{4 d_m^2} \;,\\
\overline{(Z_m-d_m)(Z_n-d_n)} & = & \sigma_{mn} + \left[ \frac{1}{4 d_m d_n} -
\frac{\sigma_{mn}}{2 d_m^2 d_n^2}\left( d_m^2 + d_n^2 -\sigma_{mn}d_m d_n
\right)\right]\;,\\
\overline{(Z_m-d_m)^3} & = & \frac{3}{2 d_m}\;, \\
\overline{(Z_m-d_m)^2(Z_n-d_n)} & = & \frac{1+r_1^2+s_1^2}{d_n}+
\frac{\sigma_{mn}}{d_m} -\frac{1}{2 d_n} -\frac{\sigma_{mn}^2}{d_n} \;,\\
\overline{(Z_m-d_m)(Z_n-d_n)(Z_o -d_o)} & = & \frac{1}{2}\left[
-\left(\frac{\sigma_{no}}{d_m}+\frac{\sigma_{mn}}{d_o} +\frac{\sigma_{mo}}{d_n} 
\right) + 2\left( \frac{\widetilde{\sigma}_{om}\widetilde{\sigma}_{nm}}{d_m}
 - \frac{\widetilde{\sigma}_{mn}\widetilde{\sigma}_{on}}{d_n}
+\frac{\widetilde{\sigma}_{om}\widetilde{\sigma}_{on}}{d_o} \right)
\right]\;,\\
\end{eqnarray}
where,
\begin{equation}
\widetilde{\sigma}_{ij} = 
{1\over d_i d_j}\left[r(\mu_{i1}\mu_{j1}-\mu_{i2}\mu_{j2})
+ s(\mu_{i1}\mu_{j2}+\mu_{i2}\mu_{j1})\right]\;.
\end{equation}
Other moments up to the third order can be constructed from the above 
expressions by substituting appropriate indices.

Given a bank of templates and the parameters of a signal, the subset 
$\{ Z_1,\ldots,Z_m\}$ can be chosen as follows. Let the coordinate of the
signal be $\theta^\prime_s = (t_a^s, \tau_{1.5}^s, \tau_0^s)$.
 A set of templates is chosen
from the template bank which lie in a neighborhood of 
$(\tau_{1.5}^s,\tau_0^s)$, where the size of this neighborhood is adjustable.
In the rectified output of each of these templates, the sample with the largest
value of $d$ is identified. We denote the  coordinate of such a  sample by  $\theta^\prime_{\alpha,j}$,
where the first index stands for the intrinsic 
parameters~($\tau_{1.5}$ and $\tau_0$)
 of the template and the second stands for the location of the 
sample within the rectified output of this template. From Eq.~(\ref{dforS}),
it follows that the location of this sample for a given $\alpha$
 can be found by maximising
 $H(\theta^\prime_{\alpha,k},\, \theta^\prime_s)$ over the time of arrival $k$. 
 For each $\theta^\prime_{\alpha,j}$, we also choose $2 n$ neighboring 
samples in the same rectified output,
 namely, the samples $\{ \theta^\prime_{\alpha,j+k}\}$, 
where  $-n \leq k \leq n$ and 
$k \neq 0$. Finally, the set of all these samples, namely, the
set $\{\theta^\prime_{\alpha,j}\}$ and {\em for each} $(\alpha, j)$, the set 
 $\{ \theta^\prime_{\alpha,j+k}\,;\,
-n \leq k \leq n, k \neq 0 \}$, gives us the required subset of rectified
output samples. Note  that $H(\theta^\prime_1, \, \theta^\prime_2)$
  plays a central role in the determination of this subset.
In our analysis we find that for most of the cases, $n=1$ or keeping only the
two nearest neighbours to $\theta_{\alpha,j}$ is a good approximation. The
choice of the neighborhood of templates
 is intimately related to the placement of the
templates themselves and is the subject of the sections which follow. 

 The distribution of the test statistic $\Lambda$, in the 
presence of a signal, is that of the maximum of the set $\{ Z_1,\ldots ,Z_m\}$.
The joint distribution of this set was shown above
 to be well approximated by a multivariate Gaussian when the strength of the 
signal is sufficiently high. An analytical form for the distribution
of the maximum of a set of  Gaussian random variables is known only for the
bivariate case. There are some approximate methods~\cite{KS}
 for the calculation of the
distribution but these are impractical for more then four or five variates.
However, 
the distribution for a larger number of variates can be estimated
from Monte Carlo simulations. A large number of realizations are generated and
for each realization, the maximum value is recorded and finally an estimate of
of the required distribution is obtained. 

Given the covariance matrix ${\bf C}_{X}$ of a multivariate Gaussian random
 variable $\overline{X}= (X_1,\ldots,X_N)$, a realization of $\overline{X}$ can be 
generated as follows. Let $\{ \lambda_i;\, 1\leq i \leq N\}$ be the
set of  eigenvalues of ${\bf C}_X$.
 Let ${\cal E}_{\rm vec}$ be an $N\times N$ matrix whose
$i$th column is the eigenvector of ${\bf C}_X$ corresponding to $\lambda_i$.
If $\overline{W} = ( W_1,\ldots ,W_N)$ is a zero mean 
multivariate Gaussian with a covariance matrix given by
 ${\rm diag}(1,1,\ldots ,1)$, 
then
\begin{equation}
\left[ \begin{array}{c}
x_1 \\
x_2 \\
\vdots\\
x_N
\end{array} \right] = 
{\cal E}_{\rm vec} \left[
 \begin{array}{c}
\sqrt{\lambda_1}\, w_1 \\
\sqrt{\lambda_2}\, w_2 \\
\vdots\\
\sqrt{\lambda_N}\, w_N
\end{array} \right]
\label{gengauss}
\end{equation}
If $\overline{X}$ has a non-zero mean vector then this can be added to the 
RHS of the above expression. It is easy to understand the above expression
 geometrically.
In an $N$ dimensional cartesian space, realizations of $\overline{W}$ will
be distributed in a spherically symmetric ``cloud''. Multiplying the 
components of $\overline{W}$ by $\sqrt{\lambda}_i$ turns this spehrical 
distribution into an ellipsoidal one. This is the distribution expected for
realizations of $\overline{X}$ in the principal axes frame of ${\bf C}_X$. 
Finally, a rotation from the principal axes frame to the actual frame is
applied. Another method that can be used~\cite{Rubinstein} 
is to perform the Cholesky 
decomposition of ${\bf C}_X$ with the elements of one of the factors
chosen in such a way as to give the correct covariances for the components
of $\overline{X}$. We use the method of Eq.~(\ref{gengauss}) in our analysis.
 
An estimate of the distribution of $\Lambda$ can also be obtained using the kind
of Monte Carlo simulation that is conventionally used for studies of
parameter estimation accuracy. In such a method, a number of 
realizations of a  noisy detector output time series are generated. 
For each such data segment, rectified outputs are generated for a set of
 templates and the location of the maximum over the outputs is recorded. The
distributions of the coordinates of the maximum then give an estimate of 
parameter estimation accuracy. One can also record the values of
the maximum and, thus, obtain the distribution of $\Lambda$. Note that a 
distribution obtained in this way would be free of any approximations.

However, there are some limitations to this method. The first is computational.
In a typical simulation in our context, each realization of
noisy data would have $\sim 10^4$ samples (for a $1.4,\,1.4\, M_\odot$ binary),
  for the case of initial LIGO, and it would be processed through $\sim 5$ 
templates. This leads to $\sim 10^6$ floating point operations~(flop) for each
realization~\cite{flops}. A simulation with $\sim 2000$ realizations would thus
involve performing $\sim 2\times 10^{9}$  flop~(we have neglected the cost of
generating the noise realization itself).
 This is not a large requirement
computationally but when the same calculations are repeated for the advanced 
LIGO case, for signals having comparable masses, this requirement becomes $\sim
10^{12}$ flop. This is because the duration of a signal with the above masses
is $\tau_0 \sim 10^3$
 sec for the case of advanced LIGO. Even on a 300~Mflops machine~(where an
Mflop is $10^6$~flop and ``flops'' stands for flop/sec), a typical high end computing power, it would take $\sim 1$ hr to complete the simulation.
 In our analysis, detection probabilities 
would be required for various configurations~($\sim 10^2$~)
 of template placement and would, thus, be quite impractical to compute using
 this method. 
There is also a more fundamental limitation. Pseudo-random number generators
have, in general, a finite period~\cite{KNUTH}. For instance, the basic
generator provided in the NAg library of numerical routines has a recommended
maximum output of $4.0 \times 10^8$ random numbers. For the advanced LIGO
case, therefore, it is actually not possible to generate more than $\sim 200$
realizations. This, of course, would lead to very poor statistics. 

On the other hand, the method represented by Eq.~(\ref{gengauss}) is 
extremely fast and since it does not depend on the signal duration, it is 
equally applicable to both the initial and advanced LIGO case. Let the 
number of samples in the set $\{ \theta_{\alpha,j} \}$ be $M$. Then the 
total number of samples in the set $\{ Z_1,\ldots ,Z_m \}$ would be $m = M\times
(2 n + 1)$. Given the covariance matrix for $\{ Z_1,\ldots ,Z_m \}$, the 
number of operations required to obtain a single realization would be 
essentially $m^2$. Typically, $M \sim 5$ and $n = 2$ which leads to 225
flop per realization, a trivial quantity computationally. Also, since for
each realization only  15 samples need to be generated, the number of trials
can be made as large as $10^7$! However, we find approximately 10,000 trials
to be sufficient for a convergence of the estimated distribution. 
Though the computational requirements for the simulation itself are small,
there is a hidden cost in this method, namely, the computation of the
 $\mu_i$, $r$ and $s$. If one were to employ FFTs for their computation, the
method would again become time consuming. However, these quantities can be 
computed quite accurately by using the stationary phase Fourier transform also.
This way, the computation of the covariance matrix also becomes quite fast. 
Typically, the whole simulation including the generation of the covariance 
matrix takes a few seconds on a 30 Mflops machine. This should be compared
 with the corresponding numbers obtained above for the conventional method.
It should be noted that this method may be used for simulations of parameter 
estimation accuracy also. Further investigations in this direction are in 
progress.

In Fig.~\ref{dpmatch}, we compare the performance of the method of multivariate Gaussian with the exact one. It can be seen that the approximate method 
becomes better as the strength  of the signal is increased. In our analysis,
a value of 0.95 for the detection probability will be taken as  fiducial and,
as can be seen from Fig.~\ref{dpmatch}, the error in the approximate method
is negligible.

We now turn to the calculation of false alarm probability. An exact 
expression for the distribution of $\Lambda$ in this case is easily obtained 
when all the rectified output samples being considered form a 
statistically independent set. In the presence of statistical correlations
between the samples, it appears that an analytical treatment is difficult.
However, it was found in MD96
that Monte Carlo estimates of false alarm probability, as a function
of threshold, could be fit almost exactly by a formula obtained by
assuming that all rectified  output samples were statistically independent
 but there were an {\em effectively} lesser number of them.
That is, if $Q_0(\eta)$ is the probability that the maximum over $N_r$
rectified output samples crosses $\eta$~(in the absence of a signal) then,
\begin{eqnarray}
Q_0(\eta) &\simeq & 1 - \exp\left[ -\epsilon N_r e^{-\eta^2/2}\right]\;,
\label{fprobapprox1}\\
	& \approx & N_r \epsilon \exp\left[-\frac{\eta^2}{2}\right]\; 
\mbox{(for $\eta \gg 1$)},\label{fprobapprox}
\end{eqnarray}
where $0<\epsilon<1$. This fit was assumed to hold for higher values of
$\eta$ also though they were beyond the range of the simulations since,
as mentioned earlier, the
period of pseudo-random number generators is finite and, consequently, it is
not possible to generate enough realizations for the estimation of
low probabilities. However,
we subsequently found that this problem
 was investigated by 
Rice~\cite{RICE}~(in 1944), though only 
for the case of a {\em single } time series. 
The formulas obtained in that work can also be interpreted in terms of an
effective number of  samples but the parameter
$\epsilon$ depends on $\eta$ and is not a constant. Specifically, $\epsilon$
approaches unity as the threshold is made higher.
Therefore, the extrapolation of the fit to Monte Carlo estimates that was made
in MD96 is not valid.   
 This does not affect the
results of MD96 significantly, however, because the quantity required in
that analysis was $\eta$ for a {\em given} false alarm and it was
shown to be highly insensitive to $\epsilon$. But it should be noted that
this implies that $Q_0(\eta)$ is affected significantly by small errors
in $\eta$ and, hence, the threshold should be recomputed once the placement of
templates is completed.

Extension of the derivation of $Q_0(\eta)$
 given by Rice~(actually, it was the {\em rate} of false alarms that was
derived) to a random {\em field} does not appear to be straightforward.  
Note that the set of rectified
outputs from a given template bank can be considered to be the samples of an
underlying 3-dimensional random field, one of the dimensions being
the time of arrival and the other two being $\tau_{1.5}$ and $\tau_0$. 
We present here a {\em qualitative} argument 
which can be extended to the case
of random fields also and show that the same conclusion regarding the 
behaviour of $\epsilon$ should hold. 

The basis of this argument is the
assumption that, in the absence of a signal, any rectified output
sample is equally likely to be the location of $\Lambda$.
 This can be expected to be true provided the random field is 
at least {\em wide sense} stationary. Since the 
input noise is assumed to be a stationary random process, it follows that
the rectified output of any one template will also be stationary. However,
the random field can be non-stationary because of
non-stationarity in $\tau_{1.5}$ and $\tau_0$ or it is genuinely stationary
 but sampled non-uniformly in $\tau_{1.5}$, $\tau_0$. For the random
field to be wide-sense stationary~\cite{Goodman},
 it is required that the correlation
of any two samples should depend on only their relative displacement and
not their locations.

In the present case,
the correlation of any two samples, $Z_1$ and $Z_2$, can be obtained
{\em exactly} in the absence of a signal. The derivation given in 
Appendix~\ref{app1} leads to the following expression for the correlation,
\begin{equation}
\overline{Z_1Z_2}  =   2 {\bf E}\left[ H(\theta^\prime_1,\,
 \theta^\prime_2)\right] - 
\left[1 -  H(\theta^\prime_1,\,\theta^\prime_2)^2\right]{ \bf K}\left[ 
H(\theta^\prime_1,\, \theta^\prime_2)\right]
 \; ,
\label{covnosig}
\end{equation}
where {\bf E} is the complete elliptic integral of the second kind and {\bf K}
is the complete elliptic integral of the first kind. Thus, 
the correlation depends
only on $H(\theta^\prime_1,\,\theta^\prime_2)$.  
Therefore, if $H(\theta^\prime_1,\, \theta^\prime_2)$
 were location independent, that is 
if it were dependent only on the difference
$\Delta \theta^\prime$ between $\theta^\prime_1$ and $\theta^\prime_2$, 
then the rectified output
field would be at least {\em wide sense stationary}. 

As discussed below,
$H(\theta^\prime_1,\, \theta^\prime_2)$ is 
location independent
in the case of Newtonian and ${\rm post}^{1}$-Newtonian wave
 forms~($\theta^\prime$ is understood to be a different set of parameters for
these wave forms) but
not in the case of 
${\rm post}^{1.5}$-Newtonian wave form. However, for simplicity, consider the  
case where $H(\theta^\prime_1,\, \theta^\prime_2)$ is
 location independent. We apply our argument to a {\em single} rectified output
first.

Since we will be using an  extremely low false alarm probability in our
analysis, the threshold required will be quite high~(typically, $8 \;{\rm to}\;9\sigma$).
Suppose that $\eta$ is so high that the probability of two or more simultaneous
crossings of $\eta$, at {\em widely}  separated locations in the time 
series, is almost zero. By a wide separation we mean roughly that the locations
are not closer than the typical correlation length scale. Of course, it is
still possible for samples which are highly correlated with the one at which
$\Lambda$ occurs, to simultaneously cross $\eta$. Thus, if $Z_i$ is the sample
at which $\Lambda$ occurs~(and crosses $\eta$), then we have assumed above 
that the probability of $Z_j$ crossing $\eta$ in the {\em same } rectified 
output is zero, where $(|j - i|)\Delta t > L$ and $L$ is the FWHM~(say) of 
the autocorrelation function. We will further assume that $\eta$ is 
sufficiently high so that if one of the neighboring samples of $Z_i$
crosses $\eta$ along with $Z_i$, 
it is {\em almost always} either $Z_{i+1}$ {\em or } $Z_{i-1}$.

To compute the false alarm probability, we need to count the number of
times $\Lambda$ crosses $\eta$ in some $N$ trials for $N \rightarrow \infty$.
Under the above assumptions, the number of favorable cases can be counted
approximately as follows. First, the number of times a given sample $Z_i$
exceeds $\eta$ is counted. For $N$ sufficiently large, it is just the {\em marginal} frequency  $n$
for that sample, 
\begin{eqnarray}
n&=& N\times \int_\eta^\infty \! dz \, z \exp\left[-\frac{z^2}{2}\right]\;,
\nonumber \\
& = & N\times \exp\left[-\frac{\eta^2}{2}\right]\;. 
\end{eqnarray}
Out of these $n$ cases, some would be such in which
 either $Z_{i-1}$ or $Z_{i+1}$
also cross $\eta$ simultaneously. The number $\widetilde{n}$
of such cases in which
$Z_{i-1}$ crosses $\eta$ would be
$\widetilde{n} = n \times p(z_{i-1}>\eta\,|\, z_i>\eta) $, 
where $p(A\,|\,B)$ is the 
probability of event $A$ given that event
 $B$ has occurred. The number of cases in which $Z_{i+1}$ crosses $\eta$ 
simultaneously with $Z_i$ would also be $\widetilde{n}$. 
Now, each sample has $n$ events associated with it that 
favor $\Lambda>\eta$ but out of these, $2\widetilde{n}$ events are common
with its immediate neighbours and these common events should be counted
only once~(recall that, by assumption, the number of events with more than
two simultaneous crossings is negligible). Thus, the total number of events $n_f$ that favor a false alarm is,
\begin{eqnarray}
n_f &\simeq & N_r\times\widetilde{n} + N_r\times(n-2\widetilde{n})\;,\nonumber\\
& = & N_r [1-p(z_{i-1}>\eta\,|\, z_i>\eta)]\times 
N\exp\left[-\frac{\eta^2}{2}\right]\;,
\end{eqnarray}
where, boundary effects have been neglected. Hence, the false alarm probability
 is,
\begin{equation}
Q_0(\eta) = 
N_r [1-p(z_{i-1}>\eta\,|\, z_i>\eta)]\exp\left[-\frac{\eta^2}{2}\right]\;.
\end{equation}
A comparison of the above with Eq.~(\ref{fprobapprox}) explains 
why the latter expression produces a good fit to Monte Carlo estimates
but now, it can be seen that $\epsilon = p(z_{i-1}>\eta\,|\, z_i>\eta)$ is
not a constant as assumed in MD96 but depends on
$\eta$,
\begin{equation}
\epsilon = \frac{1}{\exp\left[-\frac{\eta^2}{2}\right]}
\int_\eta^\infty\int_\eta^\infty du dv \, P_{Z_i, Z_{i-1}}(u,v)\;,
\label{epsilon}
\end{equation}
where, $P_{Z_i, Z_{i-1}}(u,v)$ is given in Eq.~(\ref{jpfornosig}). Even for such
a simple argument, the above expression for $\epsilon$ yields values that
are of the same order as those obtained by fitting the Monte Carlo estimates.
For instance, from Eq.~(\ref{epsilon}) we get $\epsilon = 0.33$ for $\eta = 6.0$
and $r^2+s^2 = 0.9$. The typical value for $\epsilon$ that was obtained in 
MD96 was $\sim 0.7$, for a single rectified output. 

It is also clear that the assumptions made above regarding 
 simultaneous crossings of $\eta$ are not strictly necessary. The essential
point is that simultaneous crossings reduce the number of favorable events
and, since the number of such events {\em per sample} is identical for
all samples~(under the assumption of stationarity), this can be expressed
as an effective reduction in the number of samples themselves. In this way,
the extension of the above argument to a random {\em field} is obvious and
leads to the same conclusion, namely, that $\epsilon \rightarrow 1$ for $\eta \gg 1$.
In this paper, therefore, we use Eq.~(\ref{fprobapprox1}) for the false
alarm probability but with $\epsilon = 1$.

In the following, we will need to estimate the
threshold that is required to obtain a given false alarm probability. For
small values of the false alarm probability, we get from Eq.~(\ref{fprobapprox}),
\begin{equation}
\eta \approx \left({\rm ln}\left[\frac{N_r^2}{Q_0^2}\right]\right)^{1/2}\;.
 \end{equation}
Let the number of templates in the template bank be $N_T$. 
Then, $N_r = N \times N_T$, where $N$ is the number of samples in a single
rectified output. Therefore,
\begin{equation}
\frac{\Delta \eta}{\eta} \approx \frac{\Delta N_T}{\eta^2 N_T}\;.
\end{equation}
This shows that the threshold is very insensitive to the number of templates
provided the false alarm is kept low. For instance, even if there is a 
relative error of 50\% in estimating $N_T$, the relative error in $\eta$
would just be $\sim 0.8\%$. As far as the detection probability of a signal
is concerned, such an error is entirely negligible. This point will be
of importance later in the paper. 
 
 \section{Placement of templates for a one-step search}
\label{tplace1}

In the previous section, we described a method for the calculation of the
detection probability of a given signal. This method consists of choosing a
small set of samples  ${\cal Z}$ from the rectified outputs of templates 
which are in some neighborhood of the signal. The set ${\cal Z}$
 is supposed to be such that the maximum, over all rectified output samples,
almost always occurs among the members of ${\cal Z}$ and
 this requires that each of them should have
 a high value of $d$~(typically, $d \geq 7.0$). Thus,
 the neighborhood of templates 
should be such that the maximum of each rectified output, in the absence of noise, be sufficiently large. 

This motivates the introduction of a quantity called
the {\em intrinsic ambiguity function} ${\cal H}(\tau_{1.5}^a, \tau_0^a ;\, \tau_{1.5}^b,\tau_0^b)$ which is defined as,
\begin{equation}
{\cal H}(\theta_a ,\, \theta_b) = 
\max_{t_a^a-t_a^b} H(\theta^\prime_a, \, \theta^\prime_b)
\; .\label{intambdef}
\end{equation}
In other words, this is the maximum value that the rectified output of a 
template $\theta_a$ will have if the input consists of only
a signal $ \theta_b$ having $S = 1$.
The role of templates and signals is, of course, interchangeable here.
 For $S \neq 1$, the maximum value will simply be
$S \, {\cal H}$. We term this value as the {\em observed strength} 
$S_{\rm obs}$ of the signal.

Clearly, the larger the ``width'' of the intrinsic ambiguity function, the 
more sparsely can templates be placed around a signal in order to obtain
the same detection probability. In this sense, the intrinsic ambiguity plays
a central role in the determination of the density of templates and thus
the computational cost of a one-step search. We  first discuss,
in the following
 section, 
 the calculation of ${\cal H}$ and some of its relevant properties.
This is followed by a discussion of template placement for a one-step search.

\subsection{The intrinsic ambiguity function}

Using the stationary phase Fourier transform given in Eq.~(\ref{statphase}), it can be shown easily that, 
\begin{eqnarray}
H(\theta^\prime_a,\,\theta^\prime_b) & = &
{1\over \beta} \left[ F_{\rm cos}^2(\theta^\prime_a,\theta^\prime_b) +  
F_{\rm sin}^2(\theta^\prime_a,\theta^\prime_b) \right]^{1/2} \;
 ,\label{hinstatp}\\
F_{\rm cos}(\theta^\prime_a,\theta^\prime_b) & = &  
\int_{f_a}^{f_c} {df\over S_n(f)} f^{-7/3} \cos[\alpha(f)] \label{fcos}\; ,\\
F_{\rm sin}(\theta^\prime_a,\theta^\prime_b) & =& -
 \int_{f_a}^{f_c} {df\over S_n(f)} f^{-7/3} \sin[\alpha(f)] \label{fsin}\;, \\ 
\alpha(f)& = &2\pi[\Delta t_a f +\sum_k \Delta \tau_k\psi_k(f) ] \; ;\; k \in [0,1,1.5],\\
\beta & = & \int_{f_a}^{f_c} {df\over S_n(f)} f^{-7/3}\;,
\end{eqnarray}
where $\Delta t_a = t_a^a-t_a^b$, $\Delta\tau_k = \tau_k^a-\tau_k^b$. The quantities $F_{\rm cos}/\beta$ and $F_{\rm sin}/\beta$ are
nothing but the quantities $r$ and $s$ defined in Eq.~(\ref{r}) and 
Eq.~(\ref{s})
 but expressed in terms of the stationary phase Fourier transform. They can,
of course, be 
calculated exactly by generating the  wave forms in time domain using
 Eq.~(\ref{tdwaveform}) and computing their correlations using FFT.
 We find that $F_{\rm cos}$ and $F_{\rm sin}$
reproduce the corresponding exact quantities 
quite faithfully for both the initial and advanced LIGO noise spectral
 densities. This also holds to a large extent when $f_c$ is replaced by
the least of the plunge cutoff frequencies corresponding to the two wave forms.
Thus, $H(\theta^\prime_a,\, \theta^\prime_b)$ calculated using Eq.~(\ref{hinstatp}) also
reproduces faithfully the corresponding exact results and this would also 
be true for ${\cal H}$ provided the location of the maximum in
 Eq.~(\ref{intambdef}) is obtained accurately.

We obtain the location of the maximum in Eq.~(\ref{intambdef}) in two steps.
First, an initial estimate for the required value of $\Delta t_a$ is obtained
 as described below. This is followed by a search for the true maximum around
this initial guess using a bracketing and golden search 
algorithm~(MNBRAK and GOLDEN in~\cite{NREC}). In order to get the initial
 estimate, we solve the
 integrals in Eq.~(\ref{fcos}) and Eq.~(\ref{fsin}) using a stationary 
phase approximation but  with
the point of stationarity chosen in
such a way as to yield the maximum value for the RHS in Eq.~(\ref{hinstatp}). Let the desired
stationary
point be $f = f_0$. Then, for fixed values of $\Delta \tau_{1.5}$
and $\Delta \tau_0$, the condition of stationarity yields  
\begin{eqnarray}
\Delta t_a & = & -(\Delta\tau_0+\Delta\tau_1-\Delta\tau_{1.5})\nonumber \\
& & +\Delta\tau_0 x^8 + \Delta\tau_1 x^6 - \Delta\tau_{1.5} x^5\;,
\label{deltata}
\end{eqnarray}
where, $x = [f_0/f_a]^{-1/3}$.
 Substituting the above back into the integrands in Eq.~({\ref{hinstatp}) and
maximising the resulting expression over $f_0$ yields the required value.
 This value of $f_0$ is obtained {\em only once} for a particular 
choice of $\Delta \tau_{1.5}$ and $\Delta \tau_0$. For any other $(\Delta \tau_{1.5}, \Delta \tau_0)$, the {\em same} value of 
$f_0$ is used to obtain an initial estimate for $\Delta t_a$  using Eq.~(\ref{deltata}). 
The algorithm given above is quite fast as compared to the exact calculation. 

 Contour plots of ${\cal H}(\theta_a,\, \theta_b)$, as a 
function of
$\theta_b$ with $\theta_a$ fixed, are shown in
 Fig.~\ref{intambfig}.  
Also shown are the eigenvectors of the {\em Hessian} of $\cal H$ which is
defined as,
\begin{equation}
{\bf H}_{ij}(\theta_a) = \frac{1}{2}\left. 
\frac{\partial^2 {\cal H}(\theta_a,\,\theta_b)}{\partial \theta_b^i\partial\theta_b^j} \right|_{\theta_b = \theta_a}\;. \label{hessian}
\end{equation}
Since ${\cal H}(\theta_a,\,\theta_a + \Delta\theta)$ 
is {\em maximum} at $\Delta\theta = 0$, the $\cal H$ surface
 is quadratic in a sufficiently small neighborhood of $\theta_a$ and,
 as shown in Fig.~\ref{intambfig}, the inner-most contours are elliptical.
The  orientation and axes lengths of such an elliptical contour
 can  be obtained in terms of the
 eigenvalues and eigenvectors of the Hessian. 
Let $\lambda_{1}(\theta_a)$, 
$\lambda_{2}(\theta_a)$ be the smaller and larger
eigenvalues of ${\bf H}(\theta_a)$ respectively and let 
$\widehat{ e}_{1 a}$ and $\widehat{e}_{2 a}$ be the corresponding
eigenvectors~(normalized to unity).
 Then, the length of the semi-minor and semi-major axes 
 of a contour at level $\epsilon$ are given by,
\begin{eqnarray}
l_1(\epsilon,\,\theta_a) = 
\frac{\sqrt{1-\epsilon}}{\sqrt{-\lambda_{1}(\theta_a) }}\;,\\
l_2(\epsilon,\,\theta_a) = 
 \frac{\sqrt{1-\epsilon}}{\sqrt{-\lambda_{2}(\theta_a)}}\;,
\end{eqnarray}
respectively 
while their orientation is given by the respective eigenvectors.
Note that since the eigenvalues are negative, $|\lambda_{1}(\theta_a)|
\geq |\lambda_{2}(\theta_a)|$. Since,
 we already
 have  a nearly accurate method for computing ${\cal H}$,
 ${\bf H}_{ij}$ can be calculated simply by approximating the derivatives
 in Eq.~(\ref{hessian}) by finite differences.
It should be noted that at a given location, the directions in which the 
contours at successively lower levels are most elongated suffer progressively 
larger rotations with respect to these eigenvectors. We call
this the {\em shear} of the contours. Typically, $\lambda_{i}(\theta_a)$
and $\widehat{e}_{i a}$ provide a good estimate of the size and orientation
of the contours for ${\cal H} \geq 0.95$.

The 
intrinsic ambiguity ${\cal H}$ is not independent of its location in 
parameter space. That is, if $\Delta\theta =
 (\Delta\tau_{1.5},\Delta\tau_0)$ be a displacement vector, 
${\cal H}(\theta_a,\,\theta_a + \Delta\theta) \neq 
{\cal H}(\theta_b,\,\theta_b + \Delta\theta)$ in general
for $\theta_a \neq \theta_b$. 
For our purpose, the most appropriate way of characterizing this location
 dependence would be to investigate the change in the dimensions of the
 inner-most contours of $\cal H$. This is because we are primarily 
interested in the detection of signals with low strengths and 
 for such signals the templates in a one step search would be placed
 closely. For instance, we find from earlier works~( MD96 and~\cite{bowen96}) that for low values of $S$, the spacing of templates is such that the signal with the lowest detection probability has ${\cal H}(\theta_t,\, 
\theta_s) \simeq 0.97$, for some template $\theta_t$.
In Figs.~\ref{inithessian},~\ref{inithessian2}
 and Figs.~\ref{advhessian},~\ref{advhessian2}, we have plotted
 $l_1(0.97,\,\theta_a)$, $l_2(0.97,\,\theta_a)$,
 the area of the ellipse $\pi\,l_1\,l_2$
and the angle between  the $\tau_{1.5}$ axis and $\widehat{e}_{1 a}$ as 
functions of $\theta_a$.
 The lowest contour level in
each plot is close to the minimum value which that quantity takes over the
whole of the space of interest.

\subsection{The geometrical configuration of a one-step template bank} 

In MD96, we had introduced a set of criteria which a template bank for a
one-step search was required to satisfy. These criteria were 
(C1)  every signal, having a strength $S$ greater than a given
 minimum strength $S_{\rm min}$, should have a detection probability greater
than a given minimum detection probability $Q_{\rm d,min}$ and  (C2)
 The false alarm should stay below a specified level $Q_{\rm 0,max}$. Throughout
the following, $Q_{\rm d,min} = 0.95$ and $Q_{\rm 0,max}$ is chosen to be 
such that the average rate of false events is 1 event/y.
Apart from the above criteria, we also demand (C3) 
that the templates be spaced as sparsely as possible so as to minimise the computational cost.

An obvious way to fix the  template placement would be to 
search through all the possible placement configurations and find the 
one satisfying the three criteria stated above. We have already introduced
formulas for detection and false alarm probabilities in Section~\ref{dist}
which can be used in checking C1 and C2. 
 However, such a blind search in configuration space is impossible to perform
in practice since the number of templates can be expected to be quite large.
Instead one can make some reasonable assumptions regarding the geometry 
of the final configuration and then proceed to perform a limited search
within that framework. We will now present an argument that suggests an
approximation to the optimum geometry of template placement in the 
${\rm post}^{1.5}$-Newtonian case. This approximation should be good 
enough for estimating the performance of a two-step hierarchical search
but a more careful analysis would be required when such a scheme is 
actually implemented.

\subsubsection{ The case of location independent $\cal H$}

 Consider, first,  a simple hypothetical situation in which 
the intrinsic ambiguity is location independent, that is,
${\cal H}(\theta_a,\, \theta_a +
 \Delta \theta) = {\cal H}(\Delta\theta)$.
Recall that the detection probability of a signal was determined by 
 a subset ${\cal Z}$ of samples
belonging to the rectified outputs of templates in some neighborhood of
the signal. Thus, the detection probability of a signal is determined 
entirely by the {\em local} distribution of templates around it. Let the coordinates of the signal be $\theta_s$ and 
that of the templates be $\{\theta_s+\Delta\theta_1,\ldots,
\theta_s+\Delta\theta_p\}$, where $P$ is the number of templates
in the neighborhood.
Also,
 the samples in ${\cal Z}$ which contribute most to the detection
probability are the ones which are located at the maxima of the rectified
outputs~(for zero noise) of these templates. 
 Let the set of such samples be ${\cal Z}^\prime = \{{\cal Z}^\prime_i\}
 \subset {\cal Z}$, $i=1,\ldots,P$. 
The remaining samples typically 
contribute only a few percent more to the detection probability. Then,
\begin{equation}
\overline{{\cal Z}^\prime_i} \simeq S\, {\cal H}(\theta_s, \, \theta_s+
\Delta\theta_i)\;.
\end{equation}
Now, if ${\cal H}$ is assumed to be location independent, 
 two different signals~(having the
same strength) would have the same detection probability provided 
the local configuration of templates $\{\Delta\theta_1,
\ldots,\Delta\theta_p\}$
 around them is the same. Strictly speaking, this statement is not true unless
$H(\theta^\prime_a,\, \theta^\prime_b)$ is also 
independent of location. This is because the covariance matrix~(determined 
by $H$) of the set of samples ${\cal Z}^\prime$ need not be location independent
 even though their mean values~(given by $\cal H$) may be so. However, we will
proceed with the assumption that 
variation in the covariance matrix is a negligible 
effect. The validity of this  assumption can be checked after the final 
results have been obtained, as we do below. 

An intrinsic ambiguity function which is location independent
is not unrealistic since the Newtonian~\cite{MD96,SD94} as well as the 
${\rm post}^1$-Newtonian~\cite{bowen96}
 wave forms are known to have such an intrinsic
ambiguity for the right choice of parameters. In fact, in these cases, the
function $H(\theta^\prime_1,\,\theta^\prime_2)$ itself is dependent
 only on $\Delta\theta$.
Thus, the detection probability of a signal in the case of Newtonian or
${\rm post}^1$-Newtonian wave forms depends {\em strictly} on 
the local configuration $\{\Delta\theta_1,\ldots,\Delta\theta_p\}$ alone.

 Now consider a configuration of templates in which the distribution
of templates in the space of interest is inhomogeneous but which is claimed
to satisfy C1, C2 and C3 for some  
$S_{\rm min}$. It then follows that the detection probability of a signal
with $S= S_{\rm min}$ in a region sparsely populated by templates is at
least $Q_{\rm d, min}$. However, this implies that a region where 
templates are spaced densely is over populated because, as shown above, the 
same sparse local distribution should suffice everywhere. Therefore, a further 
reduction in computational cost can be brought about by removing some of the
templates from the over populated region. This implies that
 an inhomogeneous distribution
cannot be an optimal one. Hence, when ${\cal H}$ is
location independent, templates should be distributed homogeneously which is
equivalent to placing them on a {\em regular} grid.  

A two dimensional regular grid is specified by a single unit cell. In the 
present paper we will assume that this unit cell is a parallelogram.
Such a unit cell is specified by
the lengths of two adjacent sides, $l_1$ and $l_2$,
 and the angles, $\alpha_1$, $\alpha_2$,
 that $l_1$ and $l_2$ make with some reference direction~(see
 Fig.~\ref{unitcellfig}). To find the optimal 
placement of templates, therefore, a search can be performed in $(l_1,l_2,
\alpha_1,\alpha_2)$ space for the unit cell having the largest area under the
constraint that the resulting grid of templates 
satisfy C1 and C2. Such a search would be computationally expensive even
with the fast methods that were introduced in the earlier sections.
 Note, however,
that   the unit cell with the largest area, for given $l_1$ and $l_2$,
is a rectangle. Hence, 
for given $l_1$ and $l_2$, a search should first be performed among all
{\em rectangular} unit cells to locate the ones that satisfy C1 and C2.
 This is equivalent to just a rotation of the
grid which involves only one of the angles and, hence, saves significantly
on computations.  
 Once the largest 
rectangular unit cell that satisfies C1 and C2 has been found, the search can then be extended to
non-rectangular unit cells with larger areas. 

 The computational cost can be reduced further
 by making an educated guess for the orientation of the rectangular
unit cell. For instance, let the contours of 
$\cal H$ be ellipsoids with the same orientation. Then it can be seen 
heuristically that the largest rectangular unit cell should be obtained 
when $l_1$ and $l_2$ are oriented along the major and minor axes, that is,
 the eigenvectors of the Hessian $\bf H$. If the
contours exhibit a shear, then the largest rectangular unit cells for
different values of $S_{\rm min}$ would, of course, be oriented differently.
In such a case also, the computation involved in the search
can be reduced by {\em starting} with an orientation given by the eigenvectors
of $\bf H$ and then searching a small range of $\alpha_1$ around this 
orientation.

In order to check that the above argument is reasonable, consider  Fig.~\ref{dpmap+intamb}. For a rectangular unit cell
 oriented along the $\tau_{1.5}$ and $\tau_0$ axes and having arbitrary
dimensions $l_1$ and $l_2$,  we
show the detection probability of signals, having the
same strength, which lie in the interior of the cell. The threshold has also
 been chosen arbitrarily and samples for the 
set $\cal Z$ were chosen from the rectified outputs of the four templates at
the vertices of the unit cell. 
Also superimposed on this detection probability map are the contours of 
$\cal H$ for one of the templates. It is clear from the figure that, as 
expected, the detection probability map closely follows the contours of the 
intrinsic ambiguity. Roughly speaking, the detection probability contours are
formed by the ``overlap'' of the $\cal H$ contours. Therefore,
it can be expected that if the unit cell were 
oriented along  the eigenvectors of
$\bf H$, then the area can be made larger, keeping the minimum detection 
probability the same, because this orientation would maximise the overlap of
the $\cal H$ contours. In Fig.~\ref{dpmap+intamb} we also show the 
detection probability map, for the same threshold as above, when the unit
cell is oriented along the eigenvectors of $\bf H$~($l_1$ along
 $\widehat{e}_{1a}$ and $l_2$ along $\widehat{e}_{2a}$) . The minimum detection
probability is much larger now which implies that the area can now
be increased further. In the following, we will restrict the parameter 
space for the unit cell to be only $(l_1,\,l_2)$ and orient the sides 
along the eigenvectors of the Hessian. That is, the sides of the unit cell
are given by $l_1 \, \widehat{e}_{1a}$ and $l_2 \, \widehat{e}_{2a}$.

\subsubsection{ The case of ${\it post}^{1.5}$-Newtonian $\cal H$}

Consider the family of 
${\rm post}^{1.5}$-Newtonian wave forms now. 
As shown in the previous section, $\cal H$ is not location independent in
this case. However, if this variation is small over the scale of few unit cells
then  
it can be expected that the optimum template placement for the 
${\rm post}^{1.5}$-Newtonian
case would also be close to a regular grid. In earlier works~(MD96 
and~\cite{bowen96}), the typical spacing of templates for low values of
$S_{\rm min}$ turned out to be such that 
for any signal, the value of $\cal H$
 was at least $\sim 0.97$ in some template. In the present case,
if templates are placed along
the eigenvectors of the Hessian, this would imply that the typical lengths
for the sides of the unit cell be $\sim 2\,l_1(0.97,\,\theta_a)$ 
and $\sim 2\,l_2(0.97,\,\theta_a)$. Thus,  the effect of the location dependence of $\cal H$ on the 
placement of templates can be studied by comparing  
 the change in $l_i$, over a distance $2\,l_i$
along $\widehat{e}_{ia}$~(that is, 
$l_i(0.97,\,\theta_a+\widehat{e}_{ia}) - l_i(0.97,\,\theta_a)$),
 with the value of $l_i$ at that point. For the
scales used in Fig.~\ref{inithessian} and Fig.~\ref{advhessian},
 a line along
any of the two eigenvectors would be nearly horizontal at any point. Therefore, 
 the change in $l_i$ along $\widehat{e}_{ia}$ can be obtained approximately by
 simply measuring the change over a line of constant $\tau_0$. 
 In Figs.~\ref{initvarintamb},~\ref{advvarintamb}, we plot the quantity
\begin{equation}
\delta_i(\tau_{1.5};\,\tau_0^0) = \left[1 - 
{l_i[0.97,\,(\tau_0^0,\,\tau_{1.5}+2\,\l_i^0)]\over l_i[0.97,\,(\tau_0^0,\,\tau_{1.5})]}\right]\times 100
\;,
\end{equation}
where $\l_i^0 $ is the value of $l_i$ at the intersection of the $\tau_0 =
\tau_0^0$ line and the boundary $AB$ or $BC$ as the case may
 be~\cite{mthdcrsection}. A similar quantity $\delta_\alpha$ can be constructed
for the angle $\alpha_1$ that the semi-minor axis makes with the $\tau_{1.5}$
axis,
\begin{equation}
\delta_\alpha =  \left[1 - 
{\alpha_1(\tau_0^0,\,\tau_{1.5}+2\,l_2^0)\over \alpha_1(\tau_0^0,\,\tau_{1.5})}\right]\times 100\;.
\end{equation}
This has also been plotted in  Figs.~\ref{initvarintamb},~\ref{advvarintamb}.

It is evident that the variation of the dimensions and 
orientations of the unit cells,
 over the scale of a single unit cell itself, is quite small over a large 
portion of the space of interest for both initial and advanced LIGO. In
general,
 this variation   becomes more rapid
 towards the high mass, or low
$\tau_0$, region. However, it is still small for advanced LIGO 
though it may have some significant effect in the case of initial LIGO. 
Thus, at least in the case of advanced LIGO, one can expect that  
 the optimum placement of templates will
be along an ``adiabatically'' changing grid over most of the space of
interest. Since the rate of detectable events is not expected to be large for
initial LIGO, we will not investigate the placement of templates for initial
LIGO any further here. Instead we will concentrate on advanced LIGO and 
assume that a quasi-regular grid will be obtained for initial LIGO also. 
 Further, in the following
analysis, we will approximate the quasi-regular grid above by 
a set of piecewise regular grids. That is, a set of patches covering the 
whole of the space of interest where the unit cells in each patch are 
identical but differ in dimension and orientation in different patches.

Though, in principle,
the unit cell in each patch can be determined by using the algorithm given 
earlier,  
 this would again be impractical because now a placement algorithm
 would have to search a two dimensional parameter space, $l_1$ and $l_2$, for
{\em each} patch. 
However, if the assumption that the detection probability
is almost completely determined by $\cal H$ alone were true, then 
the search would collapse to just two dimensions. This can be seen as
follows~(we call this assumption {\bf Ass1} for convenience in the following).
 For a small displacement $\Delta\theta = x_1\,\widehat{e}_{1a}
+ x_2\,\widehat{e}_{2a}$ at $\theta_a$,
\begin{equation}
{\cal H}(\theta_a,\,\theta_a+\Delta\theta)
 \simeq 1 - [\lambda_1(\theta_a)x_1^2  + \lambda_2(\theta_a)x_2^2]\;,
\label{tayloramb}
\end{equation}
where we have rotated the coordinate system locally so that ${\bf H}(\theta_a)$ is
diagonalized.
 Consider
a different location $\theta_b$ where $\lambda_i(\theta_b) = 
\alpha_i \lambda_i(\theta_a)$. Then, for a small displacement at
 $\theta_b$~(in a rotated coordinate system that diagonalizes 
${\bf H}(\theta_b)$) and at the same order of approximation as in
 Eq.~(\ref{tayloramb}),
\begin{equation}
{\cal H}[\theta_b,\, \theta_b+\sum\,x_i\widehat{e}_{ib}
/\sqrt{\alpha_i}\,]  = {\cal H}[\theta_a,\, \theta_a+
\sum\,x_i\widehat{e}_{ia}]\;.
\label{transymmamb}
\end{equation}
Now, suppose that the unit cell at $\theta_a$, having dimensions $l_1$
and $l_2$, satisfies C1 for a given threshold.
 This implies that every signal in the interior of
the cell has a detection probability greater than $Q_{d,\rm min}$.
Let the {\em relative} coordinates of such a signal be $\theta_s = (
\epsilon_1\,l_1,\, \epsilon_2\,l_2)$,
where $0<\epsilon_i<1$. 
If {\bf Ass1}  is true,  then the detection probability must depend only on 
${\cal H}_1 = {\cal H}(\theta_a+\theta_s,\, \theta_a)$, 
${\cal H}_2 = {\cal H}(\theta_a+\theta_s,\, \theta_a+
l_1\,\widehat{e}_{1a})$,
${\cal H}_3 = {\cal H}(\theta_a+\theta_s,\, \theta_a+
l_2\,\widehat{e}_{1a})$ and
${\cal H}_4 = {\cal H}(\theta_a+\theta_s,\, \theta_a+
l_1\,\widehat{e}_{1a}+l_2\,\widehat{e}_{2a})$~(we have neglected the
 contribution from other templates for the present but this does not affect
 the argument). It then follows from eq.~(\ref{transymmamb}) that, 
 for the same threshold, the 
detection probability of a signal with
relative coordinates $(\epsilon_1\,l_1/\sqrt{\alpha_1},\,
 \epsilon_2\,l_2/\sqrt{\alpha_2})$ at $\theta_b$ will be the same
as that of $\theta_s$ at $\theta_a$. Hence, for a given threshold, if
a unit cell with sides $l_1$, $l_2$ at $\theta_a$ satisfies C1, then so would a unit cell with sides $l_1/\sqrt{\alpha_1}$, $l_2/\sqrt{\alpha_2}$ at
$\theta_b$. 

Assume that for a given threshold, the largest unit cell that is compatible
with C1 is unique~(note that  the orientation of unit
 cells has already been fixed and only rectangular unit cells are being
 considered). We call this assumption {\bf Ass2}. Now, suppose that the optimum 
solution compatible with all the three criteria C1, C2 and C3 has been obtained
by some means. That is, the sides of the largest unit cell compatible with
C1 in {\em each} patch
as well as a common threshold have been found. From {\bf Ass2} and {\bf Ass1},
 it then follows that if the unit cell dimensions are $l_1$ and $l_2$ in any one
patch, the dimensions of a unit cell in any other patch must be 
$l_1/\sqrt{\alpha_1}$ and $l_2/\sqrt{\alpha_2}$. Hence, when searching the
$2\,N_p$ parameter space of unit cell dimensions~($N_p$ being the number of
patches), only the subspace $l_1$, $l_2$ for any one unit cell needs to be
searched. We should emphasize here that
the above argument is by no means a rigorous proof. Given the complicated 
interdependencies of various quantities~(for instance, even the number of 
patches and also the extent of a patch may depend on the dimensions of the unit 
cells), it would be difficult to cast the problem into a tractable 
mathematical form. However, we find the above argument sufficiently suggestive 
and the conclusions reached as plausible. 
The assumption {\bf Ass2} is actually not required since it can be expected that the
total number of templates, hence the threshold for a given false alarm, will
only depend on the areas of the unit cells in each patch and not on their
individual dimensions. Thus, one can always choose the optimum solution to
be the one where unit cells are scaled verisons of each other, without violating
C2 or C3.

How large can $l_1$ and $l_2$ be in the ${\rm post}^{1.5}$-Newtonian case
before the detection probability for scaled unit cells starts showing 
significant errors?
We have checked this empirically and the results are presented in Figs.~\ref{detponamb1}~(initial LIGO) and
 Figs.~\ref{detponamb2}~(advanced LIGO). In each figure we present our
results for different values of $S_{\rm min}$ which are chosen to encompass
the typical range of $S_{\rm min}$ that will be considered later. The 
detection probability that we consider throughout our analysis is 0.95.
Hence, we compute the errors in detection probability at the 0.95 level. Also,
it is not enough to compute the error for only one signal point since it
may depend on the signal location. Therefore, the maximum error among three
different signals is shown.

For each plot,
 we take 
two widely separated locations $\theta_A$ and $\theta_B$. The
x-axis and y-axis are the values of $l_1$ and $l_2$ at  $\theta_A$.
 The
corresponding quantities at $\theta_B$ are 
$l_1^\prime = l_1/\sqrt{\gamma_1}$ and $l_2^\prime = l_2/\sqrt{\gamma_2}$,
where $\gamma_i = \lambda_i(\theta_B)/\lambda_i(\theta_A)$.
At each location the detection probabilities of  three representative 
signals are 
obtained~(this anticipates the discussion of the Section~\ref{algofor1stp}),
 namely,
the signals $\theta_{1a} = \theta_a + (l_1\,\widehat{e}_{1a} +
l_2\,\widehat{e}_{2a})/2$, 
$\theta_{2a} = \theta_a + l_1\,\widehat{e}_{1a}/2$ and 
$\theta_{3a} = \theta_a + l_2\,\widehat{e}_{2a}/2$, 
where $a = A$ or
$B$. Let the threshold at which the detection probability of $\theta_{iA}$
equals 0.95 be $\eta_i$. Let the detection probability of $\theta_{iB}$ at the same threshold $\eta_i$ be $Q_{di}$. 
The quantity plotted on the $z$-axis is the maximum 
relative error among the three signals. That is,
$$
\max_i \; \left[ 1 - \frac{Q_{di}}{0.95} \right]\times 100\;.
$$
Thus, we are plotting the {\em maximum} relative error in the detection
probability~(at the 0.95 level) as a function of the unit cell dimensions.
As mentioned earlier, the typical one-step spacings that can be expected
are $2\,l_1(0.97,\,\theta_a)$ and $2\,l_2(0.97,\,\theta_a)$.
 For these values, we see from the 
figures that the typical error is $\leq 2\%$. In fact, the errors stay
small for much larger values of the unit cell dimensions. Hence, for a 
one-step template placement involving low values of $S_{\rm min}$, {\bf Ass1}
can be assumed to be valid for ${\rm post}^{1.5}$-Newtonian wave forms.

\subsection{The number of templates for a one-step search for
 ${\bf post}^{1.5}$-Newtonian wave forms}
\label{numtmp}

In order to obtain the computational cost of a one-step search as well
as the threshold for a given false alarm, the number
of templates in a grid has to be obtained. This is not a straightforward task,
however, because of the non-trivial shape of the boundary and the
variation in the area of the contours with change in location. 

The non-trivial shape of the boundary would cause some templates in any 
regular grid to fall outside the space of interest. However, as can be
seen from Figs.~(\ref{initligospcofint}),~(\ref{advligospcofint}), in which
 the unit cells are almost horizontal because of the scales used, such an 
effect would be significant only near the region around vertex $A$. In this
region, even a single unit cell may span both the boundary segments $AC$ and
$AB$. Note, however,
 that the segments $AB$, as well as $BC$, are not strict limits.
That is, astrophysically valid templates can also exist beyond them. This
is not true, however, for the segment $AC$ which is the image of the principal
 diagonal in the $(m_1,\,m_2)$ plane. Thus, although no template should be
placed on the left of $AC$, it is acceptable if some templates in a grid
 breach $AB$ or $BC$. This allows a quasi-regular grid to be placed in the
region near $A$, as shown schematically in Fig.~\ref{qrgridnearA}. This 
patch of templates can be used to cover the space of interest from $A$ till
that value of $\tau_0$ at which the width of the space of interest
 becomes comparable to the length $l_2$ of the unit cell. The number of
extra templates thus added will not be significant compared to the total 
number of templates that will be required to cover all of the space of 
interest. 

The effect of the variation in the area of the $\cal H$ contours on the
 number of
templates can be incorporated approximately as follows. Recall that in the 
previous section we showed that unit cells in different patches can be
taken as scaled versions of a standard unit cell, where the scale
factors were the ratio of corresponding eigenvalues of the Hessian. Thus,
the area of a unit cell will vary with location 
according to these scale factors only and, hence, the relative change in the
area will be the same as that in the 
area of the 0.97 contour which is shown in Fig.~\ref{inithessian2} and
Fig.~\ref{advhessian2}.
 
 Let $l_1$ and $l_2$
be the dimensions of the unit cell in that region of the space of interest
where the variation in the area is small~(say, maximum  relative change of
$\sim 15\,\%$). For initial LIGO this is roughly 
the region between vertex $A$ and
the contour at 0.002~${\rm sec}^2$ in Fig.~\ref{inithessian2}, while
 it is the area between $A$ and the contour at 0.04~${\rm sec}^2$ for the case
of advanced LIGO. Let $A_{C_1-C_2}$ be the area between contours $c_1$ and
$c_2$ and $a = l_1\,l_2$. Then,  for
the case of initial LIGO, the number of templates that lie in
$A_{0.002-0.003}$ would be approximately $A_{0.002-0.003}/(1.5\, a)$ since
the area of the contour increases by $\sim 50\,\%$ in this region. Similarly,
the number of templates in $A_{0.003-0.004}$ would be 
$\sim A_{0.003-0.004}/(2.0\,a)$ and so on. Let $\beta_{C_1-C_2} =A_{C_1-C_2}/A$,
where $A$ is the area of the {\em whole} of the space of interest~(see 
Eq.~(\ref{areaval})) and
$N_T^v$ be the number of templates in the region where the variation
in area is fast~(i.e., below the 0.002~${\rm sec}^2$ contour). Then, 
\begin{eqnarray}
N_T^v &\simeq &\left[ \frac{\beta_{0.002-0.003}}{1.5}+
\frac{\beta_{0.003-0.004}}{2.0}+\frac{\beta_{0.004-0.005}}{2.5}+\right.
\nonumber\\
& &\left.\frac{\beta_{0.005-0.006}}{3.0}+\frac{\beta_{0.006-0.007}}{3.5}+
\frac{\beta_{0.007-0.008}}{4.0}+\ldots\right] \frac{A}{a}\;,
\label{corrarea}
\end{eqnarray}
where we have not taken more terms because their corresponding areas are
negligible~(even $A_{0.007-0.008} = 0.07 A_{0.002-0.007}$).
The values of $\beta_{C_1-C_2}$ are, $\beta_{0.002-0.003} = 0.193$,
 $\beta_{0.003-0.004} = 0.090$, $\beta_{0.004-0.005} = 0.061$,
$\beta_{0.005-0.006} = 0.048$, $\beta_{0.006-0.007} = 0.041$ and 
$\beta_{0.007-0.008} = 0.031$. We call the coefficient of $A/a$ on the RHS of
Eq.~(\ref{corrarea}) as $\kappa$. For the case of initial LIGO, therefore,
$\kappa = 0.233$.
Similarly,  for the case of advanced LIGO, 
\begin{equation} 
\kappa =  \frac{\beta_{0.04-0.05}}{1.25}+\frac{\beta_{0.05-0.06}}{1.5}
+ \frac{\beta_{0.06-0.07}}{1.75}  \;,
\end{equation}
where $\beta_{0.04-0.05} = 0.183$, $\beta_{0.05-0.06} = 0.062$ and 
$\beta_{0.06-0.07} = 0.020$ which give, $\kappa = 0.199$. To understand
what these values for $\kappa$ mean, assume that the number of templates in
the remaining region of the space of interest~(that is, the region with a
slow variation) can be obtained by simply dividing its area by that of the
unit cell. Then the total number of templates $N_T^t$ would be,
\begin{equation}
 N_T^t = (1-\sum\beta_j)\frac{A}{a} + \kappa\, \frac{A}{a}\;.
\end{equation}
For initial LIGO $\sum\beta_j =0.464$, which gives
 $N_T^t \simeq 0.769 \, A/a$.
Similarly for advanced LIGO, $N_T^t \simeq 0.934 \, A/a$. This clearly shows
that the variation in the area of unit cells has a small effect in the case of
advanced LIGO. 

We combine the two approximations discussed above to give the following 
algorithm for estimating the total number of templates. 
 Recall that in the region of
large $\tau_0$, all the three quantities $l_1$, $l_2$ and $\alpha_1$ vary
quite slowly. For the purpose of counting the number of templates, therefore,
we will assume the orientation and dimensions of a unit cell to be 
constants in the regions (a) between vertex $A$ and the 0.002~${\rm sec}^2$
contour for initial LIGO and (b) between vertex $A$ and the 0.04~${\rm sec}^2$
contour for advanced LIGO. We choose an average value of $\alpha_1 = 38^\circ$
for initial LIGO and $\alpha_1 = 45^\circ$ for advanced LIGO, where $\alpha_1$
is the angle between the semi-minor axis and the $\tau_{1.5}$ axis. The final
results are quite insensitive to the choice of these angular values.
 In the first step
of the algorithm, we count the number of templates in the region near $A$
by placing unit cells as shown schematically in Fig.~\ref{qrgridnearA}. The
unit cells are ``stacked'' below each other until the length of the segment
along a $\tau_0 = {\rm const.}$ line equals $l_2$. Let this value of $\tau_0$
be $\tau_0^{eq}(l_2)$ and the number of templates thus obtained be $N_T^{eq}$. 
The area, $A_{eq}$,
between the vertex $A$ and the $\tau_0 = \tau_0^{eq}(l_2)$
line is then found. The total number of templates is then obtained as,
\begin{equation}
N_T^t = N_T^{eq} + \left[1-\frac{A_{eq}}{A}-\sum \beta_j\right]\frac{A}{a}+
\kappa\,\frac{A}{a}\;.
\end{equation}
 The output of this algorithm is shown in 
Fig.~\ref{tmpcountinit} and Fig.\ref{tmpcountadv}, where we have also shown
the values obtained if the number of templates is estimated simply as $A/a$.
Again, it can be seen that the effect of variation in $\cal H$ contours is small for 
advanced LIGO.

Almost all the problems associated with the non-trivial shape of the
boundary of the space of interest can be 
eliminated if instead of the segment $AB$, a 
rectangular corner $ADB$ were used. That is, $D$ has the abscissa of $B$ and
the ordinate of $A$. However, the region between the segment $AB$, $AD$ and
$DB$ is then mapped, in the $(m_1,\,m_2)$ plane, onto a negligibly small
area. Thus, although the 
number of templates will increase significantly, not much will be gained in
terms of the range of detectable binary systems.

The boundary can also be made simple by going over to a different set of 
 parameters, like the masses $(m_1,\, m_2)$. But we found that 
in such cases the intrinsic ambiguity function shows excessive location
dependence. However, there may exist a coordinate system in which both the
boundary of the space of interest is simple and the intrinsic ambiguity
function does not show much variation. This approach needs to be explored
 more thoroughly. The problems with the counting of templates, discussed
 above, are also present for the coordinates used in~\cite{bowen96,apost96}. 

To summarize, the following conclusions emerge from the discussion presented
so far. (i) In a case where the intrinsic ambiguity is location independent~(as
happens, for instance, in the Newtonian and ${\rm post}^1$-Newtonian case), 
templates should be placed on a regular grid~(neglecting boundary effects). (ii)
The unit cell of the grid should have the largest area~(in order to satisfy C3)
while satisfying C1 and C2. Assuming that the unit cell is a parallelogram, we gave a practical algorithm to find the
four parameters $(l_1,l_2,\alpha_1,\alpha_2)$ of this optimum unit cell. (iii)
The location dependence of $\cal H$ in the ${\rm post}^{1.5}$-Newtonian
case is quite weak over most of the space of interest~(at least for the case of advanced LIGO) as shown in Fig.~\ref{initvarintamb} and 
Fig.~\ref{advvarintamb}.  This implies that
the placement of templates in this case should also be on an approximately
regular grid. (iv) We can make a piecewise approximation to this grid where
each piece, or patch, is regular~(formed by translating the same unit cell). 
(v) If the detection probability of a signal were to be determined almost
completely by the intrinsic ambiguity, then only a single unit cell for
any one patch needs to be determined. We checked that this assumption
is true for the ${\rm post}^{1.5}$-Newtonian wave form.
 (vi) Since the segment
 $AB$ is not a strict boundary, it is acceptable  if some templates from a grid near  vertex $A$ breach it. We, therefore, put a quasi-regular grid of 
single unit cells ``stacked'' vertically near $A$. In the remaining region,
variation in the area of unit cells was approximately 
taken into account 
while estimating the number of unit cells.
 We emphasize here that these conclusions will
not hold for large values of $S_{\rm min}$ but only for values that are
sufficiently low so as to make the unit cells small.

\subsection{Algorithm for the determination of the optimum unit cell}
\label{algofor1stp}

We now present the algorithm that we have used in this paper for the 
determination of the parameters $l_1$ and $l_2$ of the optimum
one-step unit cell. Although the algorithm that was obtained earlier is
practical enough, we can simplify it further as follows. 

First, as has already been shown earlier~(Section~\ref{dist}), the
threshold is very insensitive to the number of templates 
 when the required
 false alarm is small. 
 Since, as shown in Fig.~\ref{tmpcountinit} and
 Fig.\ref{tmpcountadv}, the relative error between the
approximate count $A/a$ and the ``exact'' count is $\sim 30\,\%$ or less,
  the threshold is affected negligibly if the approximate
count is used in its determination.

However, it must be emphasized here that the false alarm is very sensitive
to changes in the threshold and, therefore, ultimately
the threshold should be fixed very accurately. 
 For instance, let the 
number of templates be $3\times 10^5$ and the length of a data segment be
8192~sec with a sampling rate of 2048~Hz~(typical values for advanced LIGO).
 Then, for an average false alarm
rate of 1~event/y, the required threshold is $8.661$. If an error of, say,
$-\,5\,\%$ is made in the determination of this threshold, then the false
alarm rate becomes $\sim 38$~events/y while if the error made is $+\,5\,\%$,
then the false event rate falls to 0.02~events/y. The latter situation is
definitely more preferable, however, since the detection probability of 
signals will not drop too much~(typically by $5\,\%$). This in turn 
implies that an overestimation of the number of templates is better than
an underestimation and that is precisely what happens when
 the approximate count of template is used~(provided the effect of
the boundary near vertex $A$ is negligible, which should be so for small
unit cells).
 
 Secondly, 
 given a unit cell, it
is sufficient to check that signals that have the least
detection probability be detectable with a probability 
$\geq Q_{\rm d, min} = 0.95$. For a rectangular cell, there exist three 
such signals, namely, the signals at the mid-points of $l_1$ and $l_2$ and
the signal at the the centroid of the rectangle. This is also borne out
by Fig.~\ref{dpmap+intamb}~(lowermost)
 which shows that 
these three signals lie at the minima of the detection probability map. 
This is also the reason that  such a set of signals was used in 
Figs.~\ref{detponamb1},~\ref{detponamb2}.

{\em The Algorithm for one-step template placement:}\\
(i) The value of $S_{\rm min}$ is fixed. We make a few preliminary coarse
runs to find out the range of values of $S_{\rm min}$ for which the unit
cells are sufficiently small~(so that the shear is unimportant). 
 As mentioned earlier, we keep
 $Q_{\rm d, min} = 0.95$ and $Q_{\rm 0, max}$ is kept 
such that the average rate of
false events is 1/y. Let the duration of each input 
data segment be $T$ sec. Then, 
\begin{equation}
Q_{0,max} = (T-\xi_{\rm max})/( 365 \times 24 \times 3600)\;,
\end{equation} 
where $\xi_{\rm max}$ is the duration of the longest template~(see 
Section~\ref{tsstat+comp}). Recall that we denote $T-\xi_{\rm max}$ by $T_P^0$.
\\
(ii) We choose a point in the $(\tau_{1.5},\, \tau_0)$ space such that when a 
unit cell is constructed around it, all the templates lie well within the 
boundary. For instance, in the case of initial LIGO, we choose the point 
$(1.3, 50.0)$. Let this point be $\theta_a$.\\
(iii) The unit eigenvectors $\widehat{e}_{1a}$ and $\widehat{e}_{2a}$
   of ${\bf H}$ are found at $\theta_a$. We consider rectangular unit 
cells such that 
(a) $\theta_a$ is always the same vertex for all of them 
(b) the sides are $l_1 \widehat{e}_{1a}$ and $l_2 \widehat{e}_{2a}$ with
$l_1 > 0$, $l_2 > 0$.  
 The values of $l_1$ and $l_2$ are chosen to lie on 
a regular grid in $(l_1,\,l_2)$ space. Typically, we keep
(a) $l_1 \in (0.01,\,0.05)$~sec and $l_2 \in (0.05,\,0.2)$~sec for the initial LIGO (b) $l_1 \in (0.04,\,0.1)$~sec and $l_2 \in (0.2,\,0.6)$~sec for the 
advanced LIGO. The number of grid points is kept at $\sim 10\times 10$.\\
(iv) Given a point $(l_1, \, l_2)$, the threshold $\eta$ required to get a 
false alarm $Q_{\rm 0, max}$ is computed using Eq.~(\ref{fprobapprox1}). The
total number of rectified output samples is $N_r = N_T^t\times T_P^0\times 2048$,
where $T_P^0$ is the padding in the time series of the template with the
longest duration~(see Section~\ref{tsstat+comp}) and
 $N_T^t$ is the total number of templates which is taken as 
\begin{equation}
N_T^t = \frac{\mbox{Area of the space of interest $A$}}{\mbox{area of the
unit cell = $l_1 l_2$ }}\;.
\end{equation}
(v) For each unit cell we consider the signal points
 $S_1 = \theta_a+(l_1/2)\,\widehat{e}_{1a}$, $S_2 = \theta_a+
(l_2/2)\,\widehat{e}_{2a}$
 and
$S_3 = \theta_a+(l_1/2)\,\widehat{e}_{1a} + (l_2/2)\,\widehat{e}_{2a}$. Their 
respective detection probabilities $Q_{\rm d,1}$, $Q_{\rm d,2}$ and
 $Q_{\rm d, 3}$, for the threshold $\eta$, are computed using the algorithm 
based on a multivariate Gaussian joint density~(see Section~\ref{dist}). The
set $\cal Z$ is chosen from among the rectified outputs of the templates
 at $\theta_a$,
 $\theta_a+l_1\,\widehat{e}_{1a}$, $\theta_a+l_2\,\widehat{e}_{2a}$, 
$\theta_a+l_1\,\widehat{e}_{1a} + l_2\,\widehat{e}_{2a}$,
 $\theta_a+l_1\,\widehat{e}_{1a} - l_2\,\widehat{e}_{2a}$ and
$\theta_a-l_1\,\widehat{e}_{1a} + l_2\,\widehat{e}_{2a}$. The last two 
templates are included in order to take care of any contribution to the
detection probability due to the shear in $\cal H$ contours.\\
(vi) If the minimum among \{$Q_{\rm d,1}$, $Q_{\rm d,2}$, $Q_{\rm d, 3}$\}
is larger than $Q_{\rm d, min}$ then the point $(l_1,\,l_2)$ is recorded
or else not. Let the set of unit cells which qualify thus be $L$.\\
(vii) The unit cell with the largest area among $L$ is chosen as the optimum
unit cell. \\
We do not proceed further to find a larger {\em non-rectangular} unit cell 
because for the values of $S_{\rm min}$ used, there will not be much of 
an improvement.

\section{ Two step Hierarchical search}
\label{twostp}

Our scheme for a two-step hierarchical search involves the use of two 
banks of templates, in both of which we use the same family of templates.
In one of the banks, templates are spaced sparsely in~(intrinsic)
 parameter space
and this is called the {\em first stage} bank. The {\em second stage} template
bank consists of templates placed more finely. 
The detector output is first
processed through the first stage templates and the locations of those 
templates are noted in whose rectified outputs there was at least one crossing 
of a threshold $\eta_1$~(the {\em first stage threshold}). The value of
$\eta_1$ is kept sufficiently low so that, even though the templates are
sparsely spaced, all signals~(with strength greater than some $S_{\rm min}$)
can produce at least one crossing in a nearby template with a probability of
$\sim 0.95$. In the next step, for each of the
  first stage templates that produce a crossing of $\eta_1$,
the detector output is processed through a
neighborhood of second stage templates  around it.
The maximum over these second stage outputs is then compared with a {\em
second stage threshold} $\eta_2$ to check for a detection.
In this way a significant saving in computation occurs since the number of
templates used on the whole is much less than if the second stage bank alone 
were used.
 
The first stage templates cannot be spaced too coarsely, however, 
because $\eta_1$ has to be lowered and at some point the number of crossings
because of noise alone becomes large. Since each such false crossing would
involve the use of second stage templates, the computational cost starts 
rising if the first stage templates are spaced too coarsely. Thus, there is 
a non-trivial optimization problem that needs to be solved while setting up
a two-step hierarchy.

It was shown in MD96 that the correlations between templates allows a
false event~(crossing of $\eta_2$ due to noise alone) to slip through the
hierarchy in spite of the presence of two thresholds.
This is essentially because of the fact that the hierarchy is designed to
allow the easy passage of a signal and if a noise realization is such as to
produce a crossing of $\eta_2$, which is quite high~(typically $\sim 8.0$),
then it would have sufficient ``resemblance''  to some signal~(in its phase information) to allow it to pass through the hierarchy. 
 Stated in another way, this is because, for a first stage 
template and its neighborhood of second stage templates, the crossing of
 $\eta_2$ is not
 statistically independent of a crossing of $\eta_1$.    
This implies that  the second stage template bank and threshold
should be determined in the same way as a one-step bank for the
given values of $S_{\rm min}$, $Q_{\rm d, min}$ and $Q_{\rm 0, max}$. The
function of a two-step hierarchy is, therefore, limited to providing an
estimate of the location of the global maximum. For this it utilizes the
information that the occurrence of a~(high) threshold crossing must 
generate in templates that are relatively far away from it.

We give here a brief review of the algorithm used to set up a two-step
hierarchical search in the Newtonian case. Let the spacing between
consecutive templates~(in terms of $\tau_0$) in the
one-step template bank, for given $S_{\rm min}$, $Q_{\rm d, min}$ and 
$Q_{\rm 0, max}$, be $\delta_2$. Then the first stage spacing $\delta_1$ is
taken to be $\delta_1 = k \delta_2$, $ k = 2,\, 3,\ldots$. For each 
$\delta_1$, the first stage threshold $\eta_1$ was kept such that a
signal with strength $S_{\rm min}$, lying in the middle of two consecutive
templates, has a detection probability $Q_{\rm d, min}$. The average number
of false crossings among the first stage templates can then be computed 
which, given the number of second stage templates to employ around each
 first stage crossing, in turn allows the overall average computation cost 
to be calculated. The minimum of this cost was then found as a function of
 $k$.

In the present case, a similar approach can be followed to space the first 
stage templates as was done for the Newtonian case. Here, there will be
two spacings to fix, namely, along the minor  and major axes of the 
one-step unit cells, and these can be chosen as integral multiples of the
corresponding one step spacings $l_1$ and $l_2$.
 For convenience in the
following, we denote a first stage unit cell at the location $\theta_a$
  as
$U_I(k_1,k_2,\,\theta_a)$, where the sides of the unit cells have
lengths  $k_1\times l_1$ and
$k_2\times l_2$.
 There are, however, a few complications that
arise in this approach. First, if a first stage unit cell
$U_I(k_1,k_2,\,\theta_a)$
satisfies C1, it is not implied that a unit
cell $U_I(k_1,k_2,\,\theta_b)$  at a different location  
will also do so. This is because the dimensions of a first stage unit cell
would be quite large and 
Figs.~\ref{detponamb1},~\ref{detponamb2} show that the error in detection probability rises with
an increase in the dimensions. Thus the same $(k_1,\,k_2)$
at two different locations would lead to different values of $\eta_1$ since
the first stage threshold is determined by detection probability.
We take this effect into 
account as follows. For each $(k_1,\, k_2)$, we take two 
widely separated locations $\theta_a$ and $\theta_b$
 and compute the 
 thresholds $\eta_{1a}$ and $\eta_{1b}$ that are required to make both $U_I(k_1,k_2,\,\theta_a)$ and  $U_I(k_1,k_2,\,\theta_b)$
satisfy C1. We then choose the minimum among these two as the
first stage threshold $\eta_1$.

The second complication is the boundary near vertex $A$ of the space of interest
which also disallows the same $(k_1,\, k_2)$, as in the
broader parts of the space of interest,  from being used in this region. 
Recall that the one step template grid in this region was
constructed out of {\em single} unit cells ``stacked'' 
vertically~(see Fig.~\ref{qrgridnearA}). Hence, if $k_2 > 1$, 
extra second stage templates  would be required in the region to the right of
 $AB$ which would increase the 
number of one-step templates without adding significantly to the range
of binary masses being detected. However note that if
 $U_I(k_1,k_2,\,\theta_a)$, for $k_2 > 1$, satisfies C1 for some
threshold $\eta_1$, then so would $U_I(k_1,1,\,\theta_a)$, since
the templates at the vertices will now be closer to all the signals in the
cell's interior. Therefore, while calculating the number of first stage 
templates in this region, we simply divide the number of one step templates 
by $k_1$.

Finally, the number of second stage templates that would be employed per
first stage crossing will now depend on whether the crossing occurs in the
narrow region near vertex $A$ or in the broader part of the space of interest.
Let this number be $\widetilde{n}$. 
In the broader part of the space of interest, 
\begin{equation}
\widetilde{n} = 4\,(k_1-1)(k_2-1) + 2\,(k_1-1) + 2\,(k_2-1)\;.
\label{nbrhood}
\end{equation}
 Now, the minimum of the  computational cost of a two-step search occurs when
the number of false crossings in the first stage becomes $\sim 1$. But 
most of the first stage templates will be located in the broader part of the
space of interest and, hence, most of the
false crossings will also occur in this region.
 This  implies that $\widetilde{n}$  will be as given in Eq.~(\ref{nbrhood})
for most
cases. We, therefore, take $\widetilde{n}$ to be the above for all crossings.
Note that, this assumption would lead to an overestimate of the computational
requirements for the first stage and is, thus, ``safe'' in this sense. 

Let the number of false crossings for a given input 
data segment be $n_c$. Then the
total number of templates which will be employed for that data segment would
be $ n_t^{(1)} + n_c \widetilde{n} $, where  $n_t^{(1)}$ is the total number of 
first stage templates. In the presence of a signal, there would be an extra
term of $\widetilde{n}$ in the above sum but since the event rate of signals
is expected to be quite low, this term can be neglected.
If we assume that the first stage rectified output are all statistically
independent of each other, as would be the case if they are spaced widely apart,
then the average number of false crossings $n_c^{\rm av}$ would be,
\begin{equation}
n_c^{\rm av} = n_c\times Q_0(\eta_1)\;.
\end{equation}
 $Q_0(\eta_1)$ is the probability of at least one crossing of $\eta_1$ in 
a {\em single } rectified output~(see Eq.~(\ref{fprobapprox1})),
\begin{equation}
Q_0(x) = 1 - \exp[-T_P^0\nu_s\exp(-x^2/2)]\;,
\end{equation}
where $\nu_s$ is the sampling rate and 
$T_P^0$ was defined in Section~\ref{tsstat+comp} to be the padding in the  
time series of the template wave form having the longest 
duration. Note that since $\eta_1$ would be 
$\sim 6.0$, the effective statistical independence of rectified output
samples may be less and $\epsilon$ should be less than unity. However, keeping
$\epsilon = 1$ leads to an over estimation of $n_c^{\rm av}$ and, hence, an
under estimation of the computational advantage of a two step search. 

The 
{\em average } total computational  cost for a two-step search would be,
\begin{equation}
n_t^{\rm av} = n_t^{(1)} + n_c^{\rm av} \times \widetilde{n}\;.
\end{equation}
In order to compare the performance of a two-step search with the corresponding
one-step search~(that is, for the same $S_{\rm min}$, $Q_{d, \rm min}$ and
$Q_{0,\rm max}$), we use the computational powers required for implementing
the two strategies on-line~(that is, the input data should be processed
in the same time as  required in its collection). 
The number of floating point
 operations required in a one-step
search to process $T$~sec of data can be estimated as follows: (i) The
number of flop involved in the Discrete Fourier transform~(DFT)
 $\widetilde{x}$
 of the detector
output time series $\overline{x}$
would be $3\,N\,\log_2 N$ where $N = \nu_s\,T$. However, this transform needs to be
computed only once and, thus, does not contribute significantly to the total
computational cost. (ii) For each template 
location $\theta_a$, two correlations would be required, namely, with
the quadrature components $\overline{q}_0$ and $\overline{q}_{\pi /2}$. This
involves computing the product of $\widetilde{x}$ with the 
DFTs of $\overline{q}_0$ and $\overline{q}_{\pi /2}$  followed
by an inverse DFT for each of the resulting series. Hence,
 $2\,N + 6\,N\,\log_2 N$ flop will be required here. 
(iii) Each of these
transformed  series would then have to be squared and added but
only the first $T_P^0$~sec of each series is required. This, thus, leads to
$3\,T_P^0\,\nu_s$ flop. Thus, the total
number of operations, $N_{\rm flop}$ involved in a one step search is,
\begin{equation}
N_{\rm flop} = N_T^{(t)}\times(2\,N + 6\,N\,\log_2 N+  
3\,T_P^0\,\nu_s)\;,
\end{equation}
For an {\em on-line} one-step search, $N_{\rm flop}$ operations would have to be performed in $T_P^0$ sec. Thus,
\begin{equation}
C_{\rm online}^{(1)} =\frac{N_{\rm flop}}{T_P^0}\times 10^{-9} \;
 \mbox{Gflops}\;,
\end{equation}
A similar estimate for an on-line two step search leads to an {\em average} 
computational requirement of,
 \begin{equation}
C_{\rm online}^{(2)} = \frac{N_{\rm flop}^{(2)}}{T_P^0}
\times 10^{-9} \;
 \mbox{Gflops}\;
\end{equation}
where, 
\begin{equation}
N_{\rm flop}^{(2)} = n_t^{\rm av}\times(2\,N + 6\,N\,\log_2 N+  
3\,T_P^0\,\nu_s) \;.
\end{equation}
We call the quantity $C_{\rm online}^{(1)}/C_{\rm online}^{(2)}$ as the
{\em computational advantage } $C_{\rm gain}$ of a two-step search. This
is the factor by which a two-step hierarchical search would be faster than
the corresponding one-step search in an on-line detection.

We now present our results in the form of Table~\ref{initligotable},
 for initial LIGO,
 and Table~\ref{advligotable} for advanced LIGO. 
The value of $S_{\rm min}$ for each table has been
taken sufficiently low so that the resulting one-step unit cells 
obtained are small. It was shown in MD96 that, for a given number of templates,
 the one-step threshold
is almost independent of $T$ for low false alarms. This implies that the
unit cell dimensions will also be independent of $T$~(the variation of the
threshold
in the advanced LIGO case is larger but it is still negligible). Therefore, 
the values of $l_1$ and $l_2$, for the one-step unit cell, are given in the
caption of each table. These values are for a unit cell located at
 $(1.3,\, 25.0)$~sec for the case of initial LIGO and $(13.0,\,1000.0)$~sec
for the advanced LIGO. The values of the one-step threshold~(which is the
second stage threshold $\eta^{(2)}$ for the two-step search) and the total
number of one-step templates~(obtained by taking the variation of unit cell 
areas into account) is also given.

 The first column in each table is the value of 
$T$. Since the sampling rate used in our calculation is $2^{11} =2048$~Hz and
 an FFT is most efficient when the number of samples is a power of two,
we choose $T$ to be a power of 2 also. The second and third columns
are the values of $k_1$ and $k_2$ at which the average computational
cost of the two-step search is minimised. The fourth column is the
corresponding first stage threshold $\eta^{(1)}$ and the fifth and sixth
columns are the corresponding values of $n_c^{\rm av}$ and $n_t^{\rm av}$. We have
kept only the integral part of $n_c^{\rm av}$ and $n_t^{\rm av}$ and, therefore,
$n_c^{\rm av} = 0$ means that $n_c^{\rm av} \sim 1$ or less.
The seventh column is the  computational power required for an on-line
two-step search followed by the computational power required for an on-line
one-step search in the eigth column. The last column lists $C_{\rm gain}$.

Even though we have used large values of $T$, especially in 
Table~\ref{advligotable}, such values would be difficult to use in a 
practical implementation because of memory restrictions. We have used
these values only to show the existence of a minimum in the computational
power requirement as a function of $T$~\cite{Schutz89}. It should be noted here that for the
case of advanced LIGO, the storage of the pre-computed
template wave forms is also a significant problem. For instance, even if 
we consider the average duration of templates in the advanced LIGO case to
be $\sim 100$~sec, the amount of storage required for all the $\sim 6\times 10^5$
quadrature Fourier transforms would be 
$\sim 100\times 2048\times 6\times \times 8/10 = 983$~GB~(assuming that each
sample value requires 8 Bytes of storage). This is a very low bound since
the duration of a significant number of templates will be much larger.

The results obtained above can
be checked approximately as follows. The  one-step template
 placement criterion of~\cite{bowen96}
 requires the templates to be placed such that, for any
signal, ${\cal H} = 0.97$ in at least one nearby template. Then the number of
one-step templates $N_T^t$
 would be the area $A$ of the space of interest divided 
by the area of the 0.97 contour. For advanced LIGO, $N_T^t = 20389.5/0.04 
\simeq 509739$. Thus, the 
threshold $\eta^{2}$ required, for a false alarm rate of 1 false event/y,
 would be $\eta^{(2)} = 8.722 $ for $T = 8192.0$~sec.
For the detection probability formula used in MD96, it was found
that the minimum {\em observed} strength required for a signal so that
 its detection probability be 0.95 is $S_{\rm obs} \approx \eta^{(2)}+0.67$.
 The actual
strength should, therefore, be $S_{\rm min} = S_{\rm obs}/0.97 = 9.682$.
 Roughly speaking, the decrease in $n_t^{\rm av}$,
with an increase in $k_1$ and $k_2$, is halted when $n_c^{\rm av}$ becomes
of order unity. Assuming that the number of first stage templates that is
 finally obtained is $\sim 10^4$, it would imply that,
for the above value of $T$, $\eta^{(1)}
\approx 7.026$.  
For a detection probability of 0.95 in the first stage, therefore,
the value of $S_{\rm obs}=S_{\rm min} {\cal H}^\prime$ should be 7.697,
 where ${\cal H}^\prime$
 is the value of the intrinsic ambiguity
in the middle of the sides of a first stage unit cell, i.e., ${\cal H}^\prime =
{\cal H}(\theta_a, \theta_a + k_i l_i \widehat{ e}_{ia}/2)$. 
The quantity
$k_i$  can then be calculated  as the ratio of
the dimension of the ${\cal H}^\prime$ contour
along $\widehat{e}_{ia}$ to $l_i(0.97,\theta_a)$. From the above, ${\cal H}^\prime = 0.79$ which gives
$k_1 = 7.67 $, $k_2 = 4.35$~(we have allowed $k_i$ to be non-integral here).
These values  are about the same as those in Tables~\ref{advligotable}.
 However, this approximation is crude in many ways and 
can only serve as an indicator for the kind of values one may get for $k_i$.

The savings in computational requirements achieved by a two-step search can be
more than what is obtained here if the first stage template grid is rotated
relative to the second stage grid. This is because of the shear of the contours.
In the argument given above, the quantities $k_1$ and $k_2$ were obtained 
as the ratios of $l_i(0.97,\,\theta_a)$ and the corresponding dimension of
the lower level contour ${\cal H}^\prime$. However, the direction in which the
${\cal H}^\prime$ contour is most elongated is {\em different} from  that of the
eigenvectors $\widehat{e}_{ia}$. If the first stage grid were oriented 
along the direction of maximal elongation of ${\cal H}^\prime$, the first 
stage unit cell may turn out to be larger. However, the calculation of the
number of first stage templates as well as the number, $\widetilde{n}$,
 of second stage templates would be more involved in such a case. We postpone
an investigation of this problem to a later work.  

\section{Conclusions}
\label{conclude}

We have investigated the performance of a two step hierarchical search 
 for the detection of gravitational wave signals emitted during the inspiral
of a compact binary. This work extends the investigations of MD96~\cite{MD96} 
to the more realistic case of zero spin ${\rm post}^{1.5}$-Newtonian template
and signal wave forms. 

As in MD96, we find that  a two-step search brings
about a significant reduction in computational requirements. For the case
of (i) initial LIGO noise p.s.d., a two-step search is $\sim 27.0$ times
faster than the corresponding one-step search and (ii) for the
 advanced LIGO noise p.s.d.,  a two-step search is $\sim 23.0$ times
faster than the corresponding one-step search. The range used for the 
masses $m_1$ and $m_2$ is $0.5\leq m_1\leq 30.0\, M_\odot$, $0.5 \leq m_2 \leq 30.0\, M_\odot$. 

In the analysis of MD96, the dominant problem was the calculation of 
detection probability in the presence of strong statistical correlations 
between rectified output samples. A solution to this problem was found in this
paper in the form of a semi-analytic method that reproduces the exact 
Monte Carlo estimates quite well. It is also shown here that statistical 
correlations are unimportant for the calculation of false alarm probability
when the threshold is kept sufficiently high. Therefore, the 
{\em effective} sampling rate used in MD96 is not required. 

Though the issues of detection and false alarm probabilities have been
addressed satisfactorily here, some new problems crop up in the present
analysis. Namely, the (i) location dependence of the {\em intrinsic ambiguity}
function and (ii) the non-trivial shape of the boundary of the space of
interest. Both these problems were dealt with by making some approximations.
The location dependence of the intrinsic ambiguity function seems weak
enough, at least in the case of advanced LIGO, for us to assume that the
grid of one-step templates will be an ``adiabatically'' changing regular
grid. This allows us to approximately take the effect of variations in the
 area of the contours into account. The non-trivial boundary has a significant
effect only near one of the vertices~(vertex $A$ of
 Fig.~\ref{initligospcofint}). We take this effect into account by placing
a single ``stack'' of unit cells in this region. 

The results of this paper show that the use of hierarchical methods of 
detection can be very useful for the case of coalescing binary signals and
provide a strong motivation for more detailed investigations. 
Such methods would be indispensable if the number of signal parameters 
required becomes large. For instance, if the orbital and total angular 
momenta of the binary are misaligned, there would be  significant modulations
of the phase and amplitude which can reduce the signal to noise ratio if
these effects are neglected in the template family. For such signals, a 
template family with a larger number of parameters may be required.

Many other hierarchical strategies are also conceivable and it remains to 
be seen whether they can be more effective than the two-step search analysed
here. For instance, one obvious strategy is to use a lower order template 
family as the first stage of the search and use the true wave forms, having
a larger number of parameters, as the second stage. It is not enough, though,
to only provide estimates of their performance since at some stage such 
strategies need to be implemented in practice and, as seen in
this paper, the details of the implementation can also be an involved issue. 
Also, the robustness of the placement configuration against changes in the
noise power spectral density needs to be investigated. The efficacy of
hierarchical methods~(not necessarily a two-step search)
 should also be investigated for the detection of 
continuous wave sources where the estimated computational requirements are
extremely large and far beyond presently available computing power. Further
investigations in this direction are in progress.

\section*{Acknowledgements}

I thank Prof.~S.~V.~Dhurandhar for many helpful discussions. I acknowledge 
the support provided by the Council of Scientific and Industrial Research~(CSIR)
 of India. 

\begin{appendix}
\section{ The bivariate probability density $P_{Z_1,Z_2}$ and
 $\overline{Z_1\,Z_2}$}
\label{app1}
Here, we outline the steps in the derivation of Eq.~(\ref{covnosig}). The
algebraic manipulations were performed using $\scriptstyle MATHEMATICA$.
First, the general expression for the joint bivariate probability density 
is derived without assuming the mean values of the Gaussian components 
to be zero.
Let the bivariate cumulative distribution function of 
$Z_1 = [X_1^2+X_2^2]^{1/2}$ and $Z_2 = [Y_1^2+Y_2^2]^{1/2}$ be
 $F_{Z_1,Z_2}(z_1, z_2)$, where $(X_1,X_2,Y_1,Y_2)$ is a set of jointly Gaussian random
variables with a 
covariance matrix give in Eq.~(\ref{covmat}) and mean values 
$\overline{X}_1 = \mu_1$, $\overline{X}_2 = \mu_2$, $\overline{Y}_1 = \nu_1$,
$\overline{Y}_2 = \nu_2$.
Changing the variables of integration to $X_1 = R\cos\phi$, $X_2 = R\sin\phi$,
 $Y_1 = Q\cos\psi$ and $Y_2 = Q\sin\psi$, we get,
\begin{eqnarray}
F_{Z_1,Z_2}(z_1, z_2) & = & {A\over 2\pi}
\, \int_0^{z_2}\!\!\!\!\!\!d Q \int_0^{2\pi}\!\!\!\!\!\!
 d\psi \int_0^{z_1}\!\!\!\!\!\! d R \int_0^{2\pi}\!\!\!\!\!\! d\phi
 R Q \exp\left[-{1\over 2}{R^2+Q^2\over (1-(r^2+s^2))}\right] \times \nonumber 
\\
& & \exp\left[ {Q\over (1-(r^2+s^2))}\left( (\nu_1-r\mu_1+s\mu_2)\cos\psi
+(\nu_2 -s\mu_1-r\mu_2) \sin\psi\right)\right] \times\nonumber\\
& & \exp\left[ {R\over (1-(r^2+s^2))}\left( (\mu_1-r\nu_1-s\nu_2)\cos\phi
+(\mu_2 +s\nu_1-r\nu_2) \sin\phi\right)\right] \times \nonumber \\
& & \exp\left[ {R Q\over (1-(r^2+s^2))} \left( r\cos\phi\cos\psi +
 r\sin\phi\sin\psi\right)\right]\;,\label{app1eq1}
\end{eqnarray}
where,
\begin{equation}
A = {1\over 2\pi\, {\rm det}[{\bf C}]^{1/2}} \exp\left[ -{1\over 1-(r^2+s^2)} \left(
{1\over 2}(\mu_1^2+\mu_2^2+\nu_1^2+\nu_2^2)-r(\mu_1\nu_1+\mu_2\nu_2) 
-s(\mu_1\nu_2-\mu_2\nu_1) \right)\right]\;.
\end{equation}
 Eq.~(\ref{app1eq1}) can be rewritten as,
\begin{eqnarray}
F_{Z_1,Z_2}(z_1, z_2) & = &  {A\over 2\pi}
\, \int_0^{z_2}\!\!\!\!\!\!d Q \int_0^{2\pi}\!\!\!\!\!\!
 d\psi \int_0^{z_1}\!\!\!\!\!\! d R \int_0^{2\pi}\!\!\!\!\!\! d\phi
 R Q \exp\left[-{1\over 2}{R^2+Q^2\over (1-(r^2+s^2))}\right]
 \exp\left[ {Q E \cos(\psi+\chi_1)\over 1-(r^2+s^2)}\right]\times \nonumber \\
& & \exp\left[{R \cos(\phi+\chi_3)\over 1 - (r^2+s^2)}\left[ (r^2+s^2) Q^2 +
2 D Q\sqrt{r^2+s^2}\cos(\psi+\chi_2) + D^2\right]^{1/2}\right] \;, \label{app1eq3}
\end{eqnarray}
where,
\begin{eqnarray}
E & = & \left[ (\nu_1-r\mu_1+s\mu_2)^2 + (\nu_2-r\mu_2-s\mu_1)^2\right]^{1/2}
\;, \\
D & = & \left[ (\mu_2-r\nu_2+s\nu_1)^2 + (\mu_1-r\nu_1-s\nu_2)^2 \right]^{1/2}
\;, \\
\chi_1 & = & \arctan\left[{r\mu_1-s\mu_2 -\nu_1\over r\mu_2+s\mu_1-\nu_2}\right]
\;, \\
\chi_2 & = & \arctan\left[{r \mu_1 -s\mu_2 -\nu_1(r^2+s^2)\over 
r\mu_2 +s\mu_1 -\nu_2(r^2+s^2)}\right]\;.
\end{eqnarray} 
The integral over $\phi$ can be performed to yield,
\begin{eqnarray}
F_{Z_1,Z_2}(z_1, z_2) & = & A\, \int_0^{z_2}\!\!\!\!\!\!d Q
 \int_0^{z_1}\!\!\!\!\!\! d R\,R Q \int_0^{2\pi}\!\!\!\!\!\! d\psi
\exp\left[-{1\over 2}{R^2+Q^2\over (1-(r^2+s^2))}\right] 
\exp\left[ {Q E \cos(\psi+\chi_1)\over 1-(r^2+s^2)}\right]\times \nonumber \\
& & I_0\left[ {R\over 1 - (r^2+s^2)}\left[ (r^2+s^2) Q^2 +2 D Q\sqrt{r^2+s^2}\cos(\psi+\chi_2) + D^2\right]^{1/2}\right] \;,
\end{eqnarray}
where, $I_0(x)$ is the modified Bessel function of the first kind of order zero.
The probability {\em density} function, $P_{Z_1, Z_2}$, can be obtained now 
as,
\begin{eqnarray}
P_{Z_1, Z_2}(u,v)& = &{\partial^2 F_{Z_1,Z_2}(u,v)\over \partial u \partial v}\; \nonumber
\\
& = &   u v A\exp\left[-{1\over 2} {u^2+v^2\over (1-(r^2+s^2))}\right]
\int_0^{2\pi}\!\!\!\!\!\! d\psi
\exp\left[ {v E \cos(\psi+\chi_1)\over 1-(r^2+s^2)}\right]\times \nonumber \\
& & I_0\left[ {u\over 1 - (r^2+s^2)}\left[ (r^2+s^2) v^2 +
2 D Q\sqrt{r^2+s^2}\cos(\psi+\chi_2) + D^2\right]^{1/2}\right]\;.
\end{eqnarray}
In the absence of a signal, $\mu_1 = \mu_2 = \nu_1 = \nu_2 = 0$ and the 
joint probability density reduces to,
\begin{equation}
P_{Z_1, Z_2}(u,v) =  \frac{u\,v}{\sqrt{{\rm det}{\bf C}}}\exp\left[-
\frac{u^2+v^2}{2(1-(r^2+s^2))}\right]\,I_0\left[
\frac{u\,v\sqrt{r^2+s^2}}{1-(r^2+s^2)}\right]\;.
\label{jpfornosig}
\end{equation}
Thus, the correlation $\overline{u\,v}$ can be obtained as,
\begin{equation}
\overline{u\,v} = \int_0^\infty\int_0^\infty\! du\,dv\, 
\frac{u^2v^2}{1-(r^2+s^2)}
\exp\left[-
\frac{u^2+v^2}{2(1-(r^2+s^2))}\right]\,I_0\left[
\frac{u\,v\sqrt{r^2+s^2}}{1-(r^2+s^2)}\right]\;.
\end{equation}
The above double integral is solved in~\cite{integral} from which we get,
\begin{equation}
\overline{u\,v} = 2 {\bf E}\left[\sqrt{r^2+s^2}\right] - 
\left[1 -  (r^2+s^2)\right]{ \bf K}\left[ \sqrt{r^2+s^2}\right]\;,
\end{equation} 
where
 {\bf E} is the complete elliptic integral of the second kind and {\bf K}
is the complete elliptic integral of the first kind.
\end{appendix}

\newpage

\begin{table}
\caption{ Minimum $C_{on-line}^{(2)}$ as a function of $T$ for : 
$S_{min} = 9.98$,
 $ \xi_{max} = 140.482$~sec, $Q_{d,min} = 0.95$,
$\eta^{(2)} = 8.314$, $l_1 = 0.022$~sec, $l_2 = 0.144$~sec, $N_T^t = 13279$.
\label{initligotable}}
\begin{tabular}{lllllllll} 
$T$(sec) & $k_1$ & $k_2$ &$\eta^{(1)}$& $n_c^{\rm av}$ & $n_t^{\rm av}$ &
$C_{\rm online}^{(2)}(Gflops)$ & $C_{\rm online}^{(1)}(Gflops)$&$C_{\rm gain}$\\ \hline
256.0 & 8 & 9 & 6.056 & 0 & 360 & 0.192 & 7.07 & 36.82\\
512.0 & 8 & 6 & 6.283 & 0 & 441 & 0.155 & 4.65 & 30.00\\
1024.0 & 8 & 5 & 6.484 & 0 & 490 & 0.152 & 4.12 & 27.11\\
2048.0 & 8 & 4 & 6.649 & 0 & 620 & 0.187 & 3.99 & 21.34\\
4096.0 & 8 & 4 & 6.649 & 1 & 682 & 0.207 & 4.02 & 19.42\\
8192.0 & 8 & 3 & 6.866 & 0 & 733 & 0.228 & 4.12 & 18.07
\end{tabular}
\end{table}

\begin{table}
\caption{ Minimum $C_{on-line}^{(2)}$ as a function of $T$ for : 
$S_{min} = 10.34$,
 $ \xi_{max} = 5621.51$~sec, $Q_{d,min} = 0.95$,
$\eta^{(2)} = 8.658$, $l_1 = 0.116$~sec, $l_2 = 0.560$~sec, $N_T^t = 300796$.
\label{advligotable}}
\begin{tabular}{lllllllll} 
$T$(sec) & $k_1$ & $k_2$ &$\eta^{(1)}$& $n_c^{\rm av}$ & $n_t^{\rm av}$ &
$C_{\rm online}^{(2)}(Gflops)$ & $C_{\rm online}^{(1)}(Gflops)$&$C_{\rm gain}$\\ \hline
8192.0 & 5 & 9 & 6.649 & 11 & 10188 & 9.771 & 288.48& 29.52\\
16384.0 & 4 & 7 & 7.002 & 6 & 13490 & 6.476 & 144.39& 22.30\\
32768.0 & 4 & 7 & 7.002 & 16 & 14390 & 5.709 & 119.34&  20.90\\
65536.0 & 4 & 7 & 7.002 & 35 & 16100 & 6.014 & 112.36& 18.68\\
131072.0 & 4 & 6 & 7.060 & 56 & 18935 & 7.004 & 111.27& 15.89
\end{tabular}
\end{table}

\newpage

\begin{figure} \caption{
The {\em space of interest} for the case of
initial LIGO noise power spectral density~($f_a = 40.0$~Hz). The vertices of
the space of interest
correspond to the following points in the $(m_1,\,m_2)$ plane : 
$A$ corresponds to $(0.5,\,0.5)\,M_\odot$, $B$ to $(30.0,\,0.5)\,M_\odot$
and $C$ to $(30.0,\,30.0)\,M_\odot$. 
} \label{initligospcofint}  \end{figure}

\begin{figure} \caption{
The {\em space of interest} for the case of
advanced LIGO noise power spectral density~($f_a = 10.0$~Hz).
The vertices of
the space of interest
correspond to the following points in the $(m_1,\,m_2)$ plane : 
$A$ corresponds to $(0.5,\,0.5)\,M_\odot$, $B$ to $(30.0,\,0.5)\,M_\odot$
and $C$ to $(30.0,\,30.0)\,M_\odot$.
} \label{advligospcofint}  \end{figure}

\begin{figure} \caption{
 The relative error in detection probability as 
obtained using the multivariate
Gaussian approximation and as obtained by performing an exact simulation.
Each figure shows the relative error for three values of signal strength,
$S = 8.0\;{\rm (solid)},\, 9.0\;{\rm (dotted)},\, 10.0\;{\rm (dashed)}$. As expected, the error decreases for larger 
signal strengths. The top left figure shows the locations of the 
templates~(crosses),
 used in the calculation of the detection probabilities, and the 
  signal locations~(filled
circles). The basic unit cell which is composed of templates \# 1, 2, 3 and 4,
 is oriented along the eigenvectors of the Hessian at the location of template
\# 1.
Templates \# 5 and \# 6  are included in the calculation to take into
account any possible contribution that
they may provide, because of the shear of contours,
 to the detection probabilities of signal \# 3 and \# 2 respectively. 
} \label{dpmatch}  \end{figure}

\begin{figure} \caption{
The contours of the intrinsic ambiguity function ${\cal H}(\theta_a,\,
\theta_b)$ for initial LIGO. In this figure, $\theta_a$ is kept fixed at
$\theta_a = (1.3,\,25.0)$~sec and $\theta_b$ is varied. Also
shown are the semi-minor and semi-major axes of the 0.97 contour as 
calculated from the Hessian ${\bf H}(\theta_a)$. The axes do not
appear at a right angle to one another because the axes scales are different. 
} \label{intambfig}  \end{figure}

\begin{figure} \caption{The upper figure shows
the contours of $\l_2(0.97,\,\theta_a)$ and the lower figure shows
the contours of $\l_1(0.97,\,\theta_a)$, on the space of interest for initial 
LIGO.
} \label{inithessian}  \end{figure}

\begin{figure}
\caption{The upper figure shows
the area, $\pi\,l_1(0.97,\theta_a)\, l_2(0.97,\theta_a)$
of the 0.97 contour of the intrinsic ambiguity function
$\cal H$ and the lower figure shows
the contours of $\alpha_1$, the angle~(in degrees) between
$\widehat{e}_{1a}$ and the $\tau_{1.5}$ axis, on the space of
 interest for initial LIGO.
}\label{inithessian2}  
\end{figure}

\begin{figure} \caption{ The upper figure shows
the contours of $\l_2(0.97,\,\theta_a)$ and the lower figure shows
the contours of $\l_1(0.97,\,\theta_a)$, on the space of interest for 
advanced 
LIGO.
} \label{advhessian}  \end{figure}

\begin{figure}
\caption{The upper figure shows
the area, $\pi\,l_1(0.97,\theta_a)\, l_2(0.97,\theta_a)$
of the 0.97 contour of the intrinsic ambiguity function
$\cal H$ and the lower figure shows 
 contours of $\alpha_1$, the angle~(in degrees) between
$\widehat{e}_{1a}$ and the $\tau_{1.5}$ axis, on the space of
 interest for advanced LIGO.
}\label{advhessian2}  
\end{figure}

\begin{figure} \caption{
 The parameters of a unit cell which is a parallelogram. Whenever the 
sides of the unit cell are assumed to be along the eigenvectors of the 
Hessian, $l_1$ is taken to be along the semi-minor axis while $l_2$ is
taken along the semi-major axis.
} \label{unitcellfig}  \end{figure}

\begin{figure} \caption{ In the uppermost figure
 we show the detection probability of signals
in the interior of a rectangular unit cell which is oriented along the 
$\tau_{1.5}$~(along the x-axis in the figure) and $\tau_0$~(along the y-axis) axes with the top left vertex at $(1.1,\,25.0)$~sec.
 The length $l_1$ of the side along the
$\tau_0$ axis is 0.05~sec while the length $l_2$ of the other side is
0.150~sec. The threshold and signal strengths were chosen~(arbitrarily) as
$\eta = 8.0$ and $S = 9.0 $. In the middle figure
the contours of this detection probability map are superimposed on some of the
contours~(dashed)
 of ${\cal H}(\theta_a,\,\theta_b)$ with $\theta_a = (1.1,\,25.0)$~sec.
In the lowermost figure, we show the detection probability map for the same unit cell but now
with $l_1$ and $l_2$ oriented along the eigenvectors $\widehat{e}_{1a}$ and
$\widehat{e}_{2a}$. The threshold and signal strength are the same as in the 
figures above it. The values along the x-axis and y-axis are the serial 
numbers of the grid points.
 } \label{dpmap+intamb}  \end{figure}

\begin{figure} \caption{The quantities $\delta_1$, $\delta_2$ and
 $\delta_\alpha$~(see Eq.~(83) and Eq.~(84))
 for the case of initial LIGO. Each row of figures corresponds
to a fixed value of $\tau_0$. For (i) the first row, $\tau_0 = 60.0$~sec
(ii) the second row, $\tau_0 = 30.0$~sec and (iii) for the third row,
$\tau_0 = 10.0$~sec. The first column corresponds to $\delta_1$, the second
corresponds to $\delta_2$ and the third to $\delta_\alpha$. The x-axis
is the $\tau_{1.5}$ axis. 
 } \label{initvarintamb}  \end{figure}

\begin{figure} \caption{The quantities $\delta_1$, $\delta_2$ and
 $\delta_\alpha$~(see Eq.~(64) and Eq.~(65))
 for the case of advanced LIGO. Each row of figures corresponds
to a fixed value of $\tau_0$. For (i) the first row, $\tau_0 = 3000.0$~sec
(ii) the second row, $\tau_0 = 2000.0$~sec and (iii) for the third row,
$\tau_0 = 800.0$~sec. The first column corresponds to $\delta_1$, the second
corresponds to $\delta_2$ and the third to $\delta_\alpha$. The abscissa
is the $\tau_{1.5}$ axis. 
 } \label{advvarintamb}  \end{figure}

\begin{figure} \caption{ The relative error in detection probability for 
corresponding signals in two widely separated unit cells. Let the locations
of the unit cells be $\theta_a$ and $\theta_b$. 
In this figure, we consider the case of initial LIGO and place the top left
templates of the unit cells at 
$\theta_a = (1.3,\,50.0)$~sec and 
$\theta_b = (1.5,\,10.0)$~sec. The value used for the signal strength $S$ is
shown at the top of each plot. The effect of plunge cutoff has been incorporated
in the calculations. 
 } \label{detponamb1}  \end{figure}

\begin{figure} \caption{ The relative error in detection probability for 
corresponding signals in two widely separated unit cells. Let the locations
of the unit cells be $\theta_a$ and $\theta_b$. 
In this figure, we consider the case of advanced LIGO and place the top left
templates of the unit cells at 
$\theta_a = (13.0,\,2000.0)$~sec and 
$\theta_b = (15.0,\,400.0)$~sec. The value used for the signal strength $S$ is
shown at the top of each plot.   
The effect of plunge cutoff has been incorporated
in the calculations.
 } \label{detponamb2}  \end{figure}

\begin{figure} \caption{A schematic illustration
of a quasi-regular grid of unit cells near the vertex $A$ of the space of
interest~(initial LIGO). The lengths used for the sides of the unit cells are
$l_1 = 0.02$~sec and $l_2 = 0.120$~sec. These lengths 
have been chosen arbitrarily but represent typical values obtained in a 
one-step search.
 The boundary of the space of interest is shown by the lighter
lines. The x-axis represents $\tau_{1.5}$ and $\tau_0$ lies along the 
y-axisn~(both units are in seconds).
The top left corner of each unit cell is placed on the left most boundary
which is the image of the principal diagonal in the $(m_1,\,m_2)$ plane.  
 } \label{qrgridnearA}  \end{figure}

\begin{figure} \caption{  The number of templates for the case of
initial LIGO as a function of the unit cell dimensions $l_1$ and $l_2$.
The solid contours are obtained by using the algorithm that takes the variation
of unit cell areas into account. The dashed contours are for the values obtained
by simply dividing the area of the space of interest by $l_1\times l_2$.
 } \label{tmpcountinit}  \end{figure}

\begin{figure} \caption{ The number of templates for the case of
advanced LIGO as a function of the unit cell dimensions $l_1$ and $l_2$.
The solid contours are obtained by using the algorithm that takes the variation
of unit cell areas into account. The dashed contours are for the values obtained
by simply dividing the area of the space of interest by $l_1\times l_2$. 
 } \label{tmpcountadv} \end{figure}


\begin{references}
\bibitem{flops} Let the number of samples in each time series be
 be N and the number of templates be
M. We use the following formula for the number of flop for {\em each}
realization,
$$
\left[ 2 N + 2 \left( 3 N \log_2 N \right) + 3 N \right]\times M \;,
$$
where the first term from the left is for the computation of the products of
Fourier transforms~(the factor 2 is for the two quadrature filters), the 
second is for their inverse transforms and
 the third for squaring and adding the outputs.
\bibitem{mthdcrsection}
 The contours in Figs.~5,~6 and Fig.~7,~8 were
 obtained by mapping the
corresponding contours obtained on a regular grid in $(m_1,\,m_2)$ space.
it is not possible to use a uniform grid of points in $(\tau_{1.5},\,\tau_0)$
 space because of the non-trivial boundary.
 To obtain $\delta_i$, we interpolate
on the regular grid in $(m_1,\,m_2)$ and take cross-section of this 
interpolating function along the desired curve~(say, a constant $\tau_0$ curve).
\end{references}
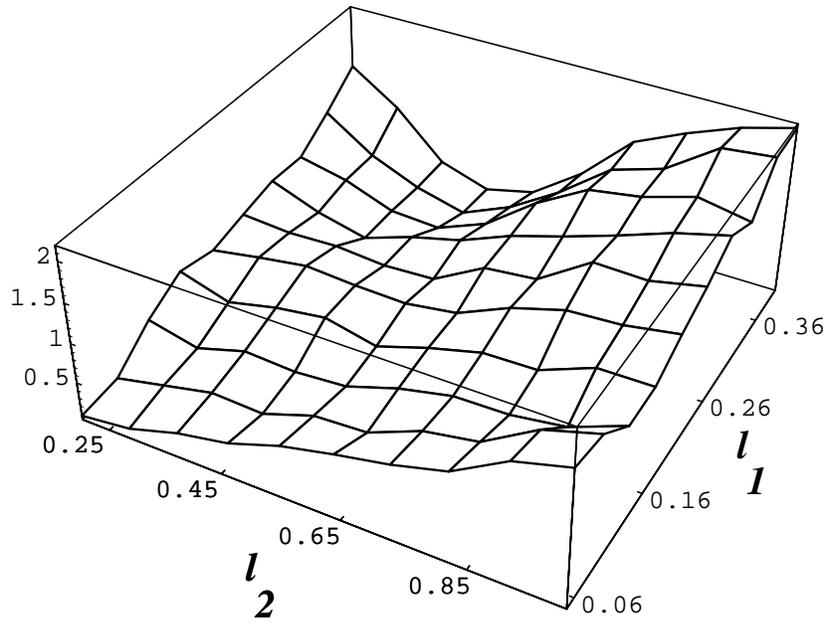
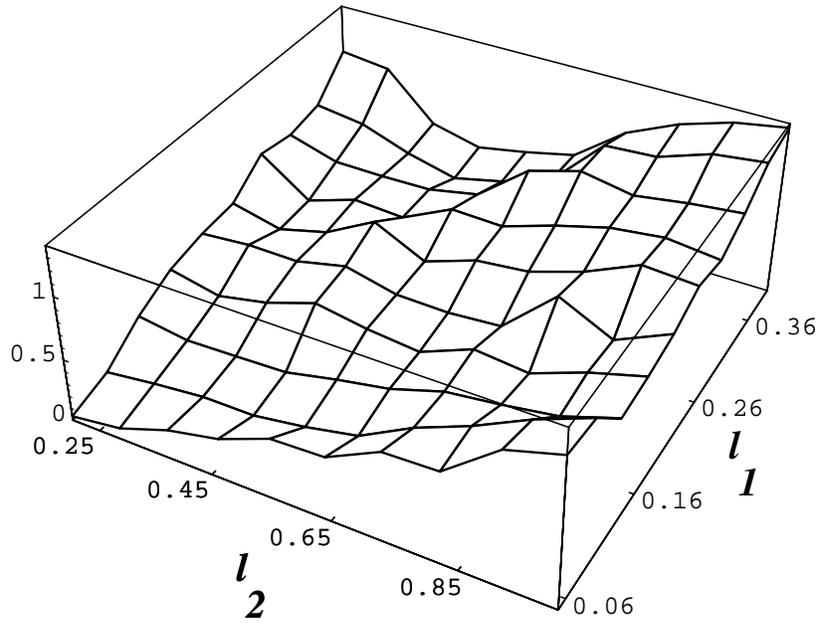
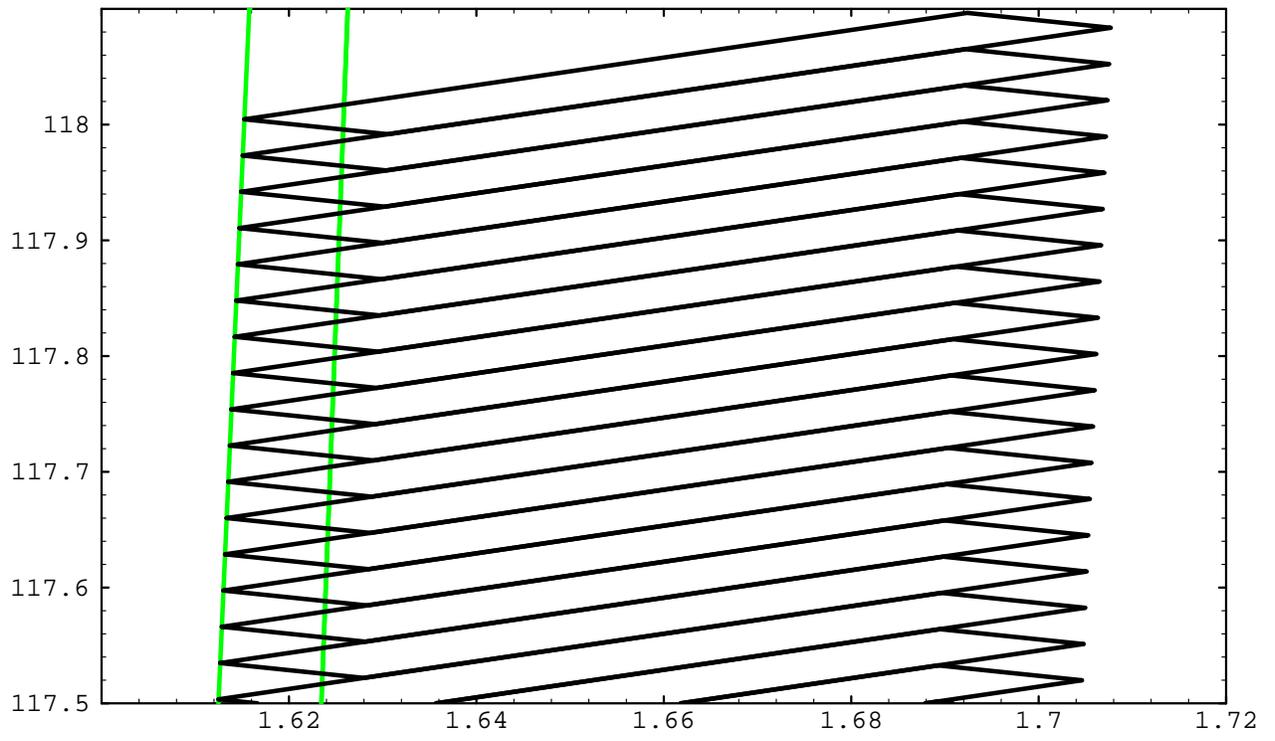
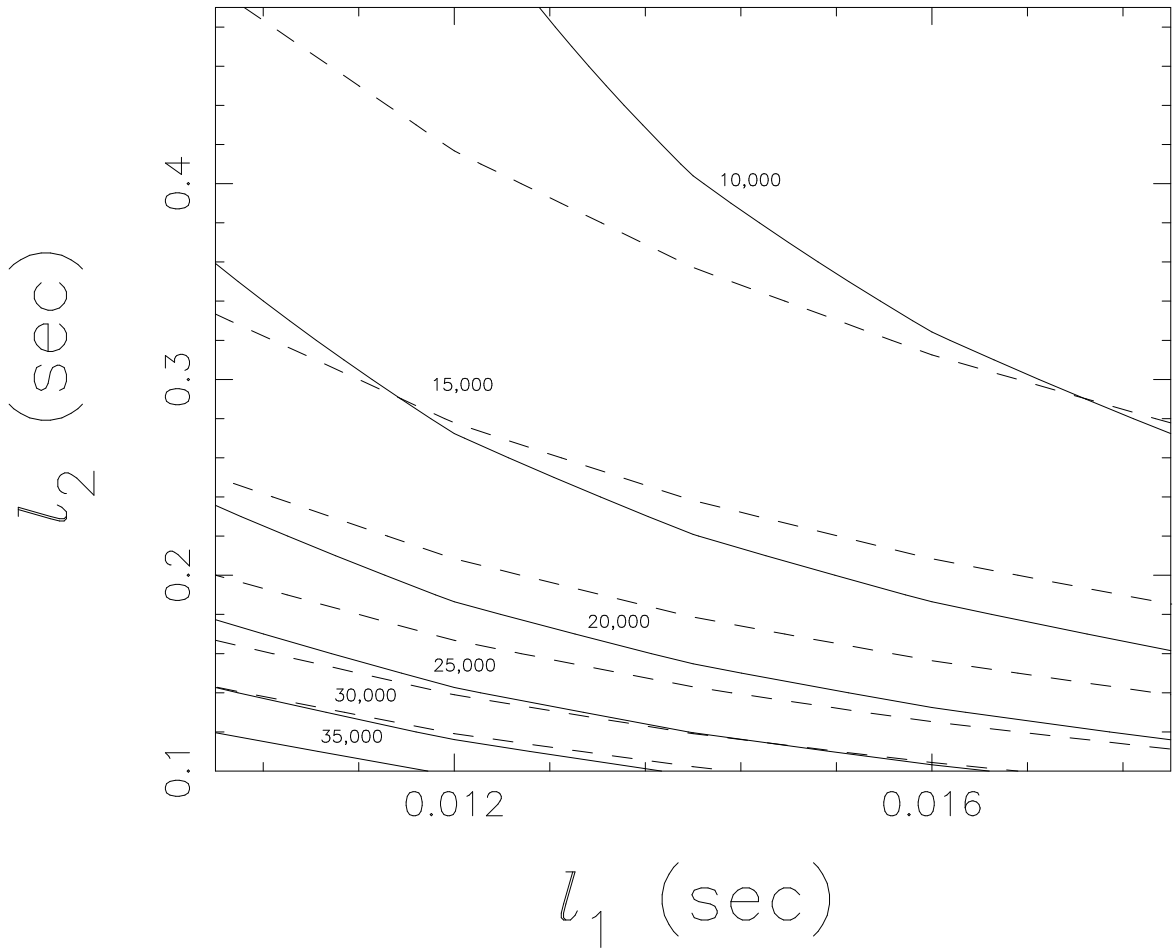
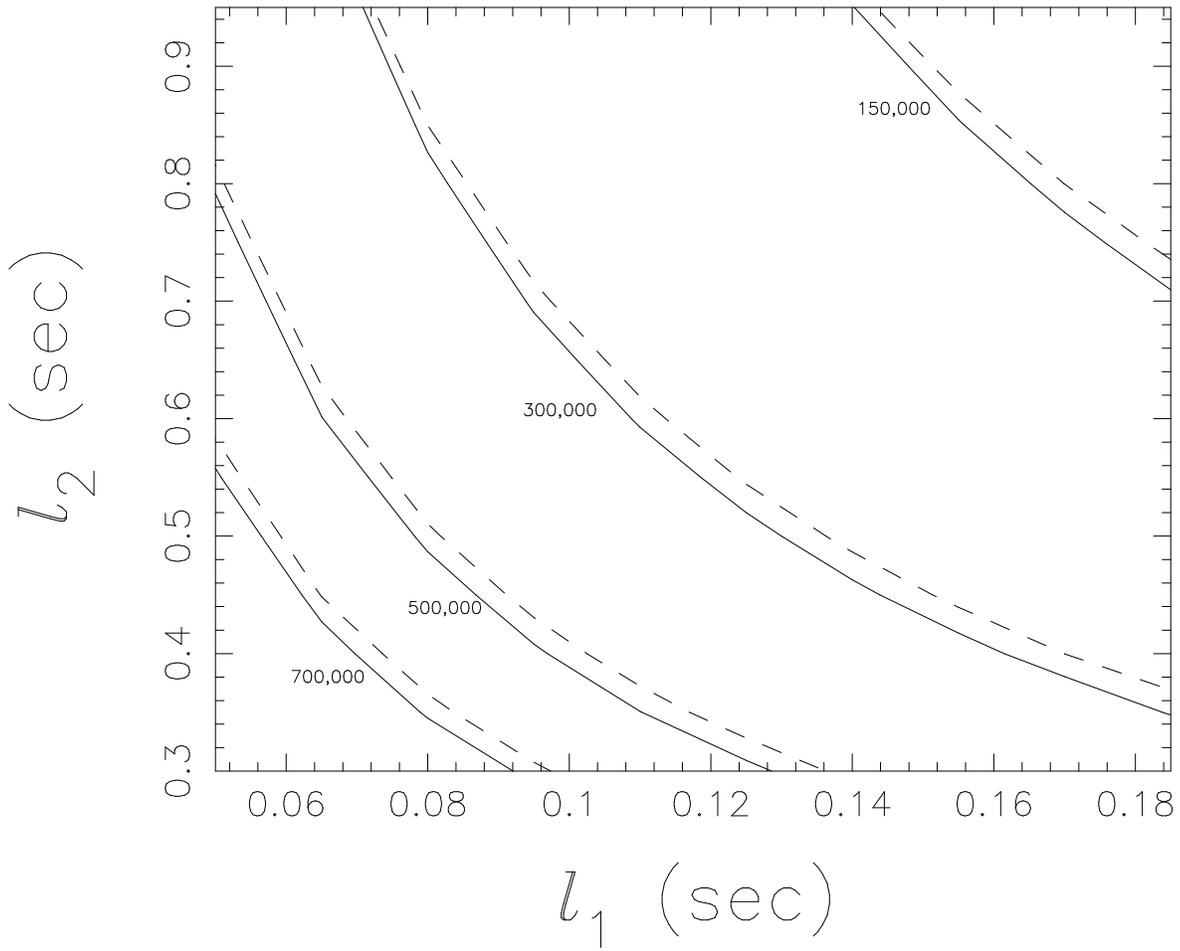
\end{document}